\documentclass[aps,notitlepage,bibnotes,longbibliography,superscriptaddress]{revtex4-1}

\usepackage{graphicx}
\usepackage{amssymb}
\usepackage{amsmath}
\usepackage{amsfonts}
\usepackage{color}
\usepackage[normalsize]{subfigure}
\usepackage{hyperref}
\usepackage{bm}
\usepackage{cleveref}
\usepackage{todonotes}

\usepackage[utf8]{inputenc} 
\usepackage[T1]{fontenc} 
\usepackage[english]{babel} 

\DeclareMathAccent{\ring}{\mathalpha}{operators}{"17}
\providecommand{\st}[1]{_{\text{#1}}}

\providecommand{\sfrac}[2]{#1/#2}

\providecommand{\ut}[1]{^{\text{#1}}}
\providecommand{\op}[1]{{\boldsymbol{#1}}}

\def\pd{\partial}

\def\im{\mathrm{i}}

\def\uv{\bv{u}}
\def\jv{\bv{j}}
\def\uv{\bv{u}}

\def\vv{\bv{v}}

\def\b0{\bv{0}}

\def\RE{\text{Re}}
\def\IM{\text{Im}}

\def\eff{\st{eff}}

\def\Bcal{\mathcal{B}}

\def\Ocal{\mathcal{O}}
\def\Scal{\mathcal{S}}
\def\gdot{{\dot{\gamma}}}

\def\d{\mathrm{d}}
\def\yield{\ut{yield}}

\def\bvisc{b}
\def\psize{a}
\def\kbT{k_B T}

\newcommand{\bitem}{\begin{itemize}}
\newcommand{\eitem}{\end{itemize}}
\newcommand{\benum}{\begin{enumerate}}
\newcommand{\eenum}{\end{enumerate}}
\newcommand{\bblock}[1]{\begin{block}{#1}}
\newcommand{\eblock}{\end{block}}
\newcommand{\bmini}[1]{\begin{minipage}{#1}}
\newcommand{\emini}{\end{minipage}}
\newcommand{\btab}[1]{\begin{tabular}{#1}}
\newcommand{\etab}{\end{tabular}}
\newcommand{\btabn}[1]{\begin{tabular}{#1}}
\newcommand{\etabn}{\end{tabular}}
\newcommand{\beq}{\begin{equation}}
\newcommand{\eeq}{\end{equation}}
\newcommand{\beqn}{\begin{equation*}}
\newcommand{\eeqn}{\end{equation*}}
\newcommand{\bmult}{\begin{multline}}
\newcommand{\emult}{\end{multline}}
\newcommand{\bsplit}{\begin{split}}
\newcommand{\esplit}{\end{split}}
\newcommand{\bmat}{\begin{pmatrix}}
\newcommand{\emat}{\end{pmatrix}}
\newcommand{\bv}[1]{\mathbf{#1}}

\begin{document}
 \title{Shear-density coupling for a compressible single-component yield-stress fluid}
 \author{Markus Gross}
\affiliation{Max-Planck-Institut f\"{u}r Intelligente Systeme, Heisenbergstra{\ss}e 3, 70569 Stuttgart, Germany}
\affiliation{IV.\ Institut f\"{u}r Theoretische Physik, Universit\"{a}t Stuttgart, Pfaffenwaldring 57, 70569 Stuttgart, Germany}
 \author{Fathollah Varnik}
 \affiliation{Interdisciplinary Centre for Advanced Materials Simulation (ICAMS), Ruhr-Universität Bochum, Universitätsstraße 150, 44801 Bochum, Germany}

\date{\today}

\begin{abstract}
Flow behavior of a single-component yield stress fluid is addressed on the hydrodynamic level. A basic ingredient of the model is a coupling between fluctuations of density and velocity gradient via a Herschel-Bulkley-type constitutive model. Focusing on the limit of low shear rates and high densities, the model approximates well---but is not limited to---gently sheared hard sphere colloidal glasses, where solvent effects are negligible. A detailed analysis of the linearized hydrodynamic equations for fluctuations and the resulting cubic dispersion relation reveals the existence of a range of densities and shear rates  with growing flow heterogeneity. In this regime, after an initial transient, the velocity and density fields monotonically reach a spatially inhomogeneous stationary profile, where regions of high shear rate and low density coexist with regions of low shear rate and high density. The steady state is thus maintained by a competition between shear-induced enhancement of density inhomogeneities and relaxation via overdamped sound waves. An analysis of the mechanical equilibrium condition provides a criterion for the existence of steady state solutions. The dynamical evolution of the system is discussed in detail for various boundary conditions, imposing either a constant velocity, shear rate, or stress at the walls.
\end{abstract}


\maketitle

\section{Introduction}

Heterogeneous flow and shear banding are ubiquitous phenomena, commonly occurring in a variety of complex fluids such as polymer solutions and worm-like micelles~\cite{Fielding2003b,Cromer2013,Moorcroft2014}, colloidal gels~\cite{Moeller2008}, hard sphere colloidal glasses~\cite{besseling_shear_2010,mandal_heterogeneous_2012} and granular media~\cite{Losert2000}. In line with this diversity of the physical systems, one encounters different underlying mechanisms as being responsible for localized flow. 
Classically, shear banding occurs in systems with a strongly shear-thinning flow curve (stress versus imposed shear rate)~\cite{Dhont2017}.
Alternatively, banding can result for a non-monotonic flow curve stemming from a shear-induced phase transformation. In this case, an instability occurs if the globally imposed shear rate lies between the two solutions corresponding to homogeneous steady flow. The system divides into two regions, each flowing with one of the stable shear rates \cite{Fielding2003b}.
For colloidal gels, on the other hand, the mechanism of shear localization is attributed to a competition between formation and growth of fractal-like clusters and its shear-induced breakage~\cite{Moeller2008}. 
 
An interesting case occurs in dense suspensions of hard sphere colloidal particles and granular materials, where the underlying flow curve is monotonic, yet the flow can develop spatio-temporal heterogeneities~\cite{besseling_shear_2010,Losert2000}.  In these ``soft glassy materials'' \cite{sollich_rheological_1998}, shear-induced rejuvenation competes with the sluggish relaxation (aging) kinetics and may lead to a heterogeneous flow in the glassy state \cite{Varnik2003,Varnik2004,Varnik2008}.

Flow localization in dense hard-sphere suspensions has been recently rationalized in terms of the so-called shear-concentration coupling (SCC) \cite{besseling_shear_2010,mandal_heterogeneous_2012}, a hydrodynamic model, first proposed in Ref.\ \cite{schmitt_shear-induced_1995}, which couples the local flow to the concentration field. This coupling is encoded in a non-Newtonian stress and in a shear-rate dependent osmotic pressure.

Within SCC, one considers a background fluid which transports---and is influenced by---a concentration field. While this picture emerges naturally in the case of polymer solutions, the role of the background fluid is less obvious in hard sphere colloidal glasses. Indeed, there is a common consensus that the effect of hydrodynamic interactions can be neglected in colloidal hard-sphere systems close to the glass transition in the low shear rate limit, which is of primary interest to the present study~\cite{phung_stokesian_1996,fuchs_schematic_2003}. Accepting this standpoint, it is tempting to fully neglect the background fluid and investigate the issue of flow heterogeneity within hydrodynamic equations of a \emph{single-component} non-Newtonian fluid. This paper presents such a study.

In Refs.\ \cite{schmitt_shear-induced_1995,besseling_shear_2010,jin_flow_2014}, the instability of a sheared colloidal suspension has been investigated based on an advection-diffusion equation for the colloid concentration $\rho$, embedded in a solvent of velocity $\uv$, 
\beq \pd_t \rho = - \nabla \cdot \jv, \qquad \jv \equiv \rho \left( \uv - \frac{1}{\zeta} \nabla \mu\right),
\label{eq_advdiff}\eeq 
where $\jv$ denotes the total particle flux, $\zeta$ is a friction coefficient, and $\mu$ is a (shear-rate dependent) generalized chemical potential \footnote{Note that $\mu$ is, in fact, not a proper chemical potential since the shear rate is a non-conservative external field. In Ref.\ \cite{jin_flow_2014} and here, the actual theoretical development does not rely on this notion but instead on a (well-defined) shear-rate dependent pressure.}.
\Cref{eq_advdiff} asserts that the total flow velocity $\jv/\rho$ of the colloidal particles consists of an imposed ``background'' flow $\uv$, onto which a contribution $-(1/\zeta) \nabla \mu$ due to the diffusive motion of the particles is superimposed. 
The flow velocity $\uv$ is assumed to be governed by the Stokes equation,
\beq 
\pd_t (\rho u_\alpha) = \pd_\beta \sigma_{\alpha\beta},
\label{eq_Stokes}
\eeq 
where $\op{\sigma}$ is the viscous stress tensor, which is typically given in terms of an expansion in gradients of $\uv$.
The Greek symbols stand for spatial directions ($\alpha,\beta \in \{x,y\}$ in the present 2D study) and Einstein's sum rule over repeated indices is used.

In Ref.\ \cite{mandal_heterogeneous_2012}, the possibility of a coupling between shear and concentration has been investigated in a system of hard spheres.
A constant kinetic temperature has been imposed by continuously rescaling the particle velocity during the simulations.
Notably, there is no background fluid in the system investigated in Ref.\ \cite{mandal_heterogeneous_2012}. Thus, it can be considered as an isothermal compressible single-component fluid, described by a continuity equation for the particle \emph{density} $\rho$ and a transport equation for the fluid momentum $\rho \uv$:
\begin{subequations}
\begin{align} 
\pd_t \rho &= -\pd_\alpha (\rho u_\alpha), \label{eq_cont}\\
\pd_t (\rho u_\alpha) &= -\pd_\beta \Pi_{\alpha\beta} + \pd_\beta \sigma_{\alpha\beta}. \label{eq_nse_mom}
\end{align}\label{eq_nse}
\end{subequations}
As in Ref.\ \cite{mandal_heterogeneous_2012}, $\op{\Pi}$ and $\op{\sigma}$ denote the reversible and the irreversible (viscous) stress tensors. In close analogy to shear concentration coupling, one postulates a coupling between fluid density and local shear rate, which we shall call ``shear-density coupling'' (SDC) in the following. As shown in \cref{sec_model}, this coupling is generated by reversible and viscous stresses being functions of the shear rate and density, respectively. 
In equilibrium, the divergence of the reversible stress tensor can be related to a chemical potential via $\pd_\beta \Pi_{\alpha\beta} = \rho \pd_\alpha \mu$. Beyond equilibrium, this relation serves as a definition of a shear-rate dependent chemical potential.

Before proceeding further with our analysis, a comment on the above equations is at order here.
The advection-diffusion equation \eqref{eq_advdiff} is central to dynamic density functional theory and widely used for the description of driven colloidal suspensions  \cite{dhont_introduction_1996, rauscher_dynamic_2007, scacchi_driven_2016}.
In these approaches, $\uv$ represents the velocity of the background fluid, which consists of an externally imposed component (e.g., shear flow) and a contribution arising from the hydrodynamic inter-particle interactions.
Notably, the dynamics of a subset of \emph{tagged} particles in a single-component fluid flowing with velocity $\uv$ is formally also described by \cref{eq_advdiff,eq_Stokes} \cite{liboff_kinetic_2003,reichl_modern_1998}.
In this case, the chemical potential and the viscosity would react only to the fluctuations of the tagged particles. 
However, the viscosity and the pressure are actually sensitive to the total density, since this quantity describes the caging and trapping responsible for the dynamic slowing down near the glass transition.

In view of these arguments on the single-component fluid nature of the problem, it appears more appropriate to analyze the hard-sphere system of Ref.\ \cite{mandal_heterogeneous_2012} in terms of the isothermal compressible fluid equations in \cref{eq_nse}, rather than an advection-diffusion equation. 
In passing, we remark that the different nature of the two models is also crucial in the case of critical phenomena: here, the advection-diffusion and momentum transport equations define the universality class of ``model H'', which primarily describes a binary fluid mixture \footnote{Model H applies to a non-isothermal single-component fluid upon identifying the order-parameter as a certain combination of the fluid and the energy density \cite{hansen_theory_2006,onuki_phase_2002}}. 
The isothermal single-component fluid, instead, is described in terms of a continuity equation and a momentum equation, giving rise to a critical behavior distinct from model H \cite{gross_critical_2012}.

\section{Model}
\label{sec_model}

\begin{figure}[t!]\centering
    \includegraphics[width=0.3\linewidth]{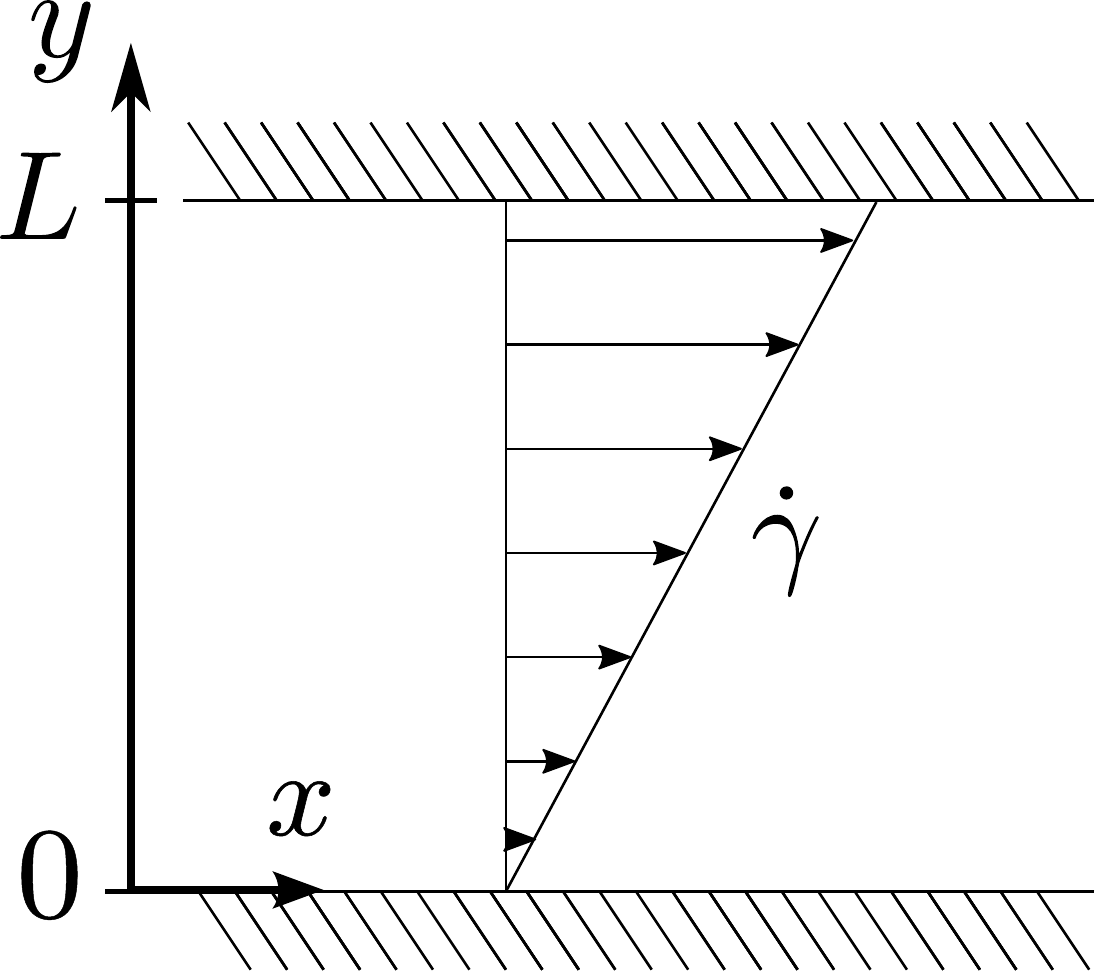} 
    \caption{Slit geometry considered in the present study. The fluid is subjected to a steady shear flow in lateral direction ($x$) with a spatially constant (background) shear rate $\gdot_0$, but can develop arbitrarily large deviations described by a local shear rate $\gdot(y,t)$. We assume the fluid to be homogeneous both in the lateral and the vorticity direction ($z$, pointing normal to the figure plane).}
    \label{fig_setup}
\end{figure}

We consider a fluid described by \cref{eq_nse}, bounded by walls at $y=0$ and $y=L$ (see \cref{fig_setup}). 
The flow is assumed to be homogeneous along the vorticity direction ($z$) as well as along the flow direction ($x$), such that generally $\pd_{x}(\cdots) = \pd_z(\cdots)=0$. 
The local shear rate is defined as
\beq \gdot(y,t) = \pd_y u_x(y,t).
\label{eq_shear_rt}\eeq 
In this and the following section, we focus on bulk dynamics, such that specification of the boundary conditions at the walls is not necessary. 
We shall therefore merely assume the presence of a constant steady background shear rate $\gdot_0$. 
(We return to the effect of boundary conditions in \cref{sec_nonlin}, where we numerically solve the Navier-Stokes equations in a finite domain.)
The pressure tensor $\op \Pi$ is isotropic, $\Pi_{\alpha\beta} = \Pi\delta_{\alpha\beta}$, where $\Pi$ denotes the scalar pressure.
Consequently, \cref{eq_nse} reduces to
\begin{subequations}\begin{align} 
\pd_t \rho &= -\pd_y (\rho u_y), \label{eq_nse_simp_rho}\\
\pd_t (\rho u_x) &= \pd_y \sigma_{xy}, \label{eq_nse_simp_jx} \\
\pd_t (\rho u_y) &= -\pd_y \Pi + \pd_y \sigma_{yy}. \label{eq_nse_simp_jy}
\end{align}\label{eq_nse_red}\end{subequations}
Analogously to Refs.\ \cite{jin_flow_2014,dhont_constitutive_1999,olmsted_phase_1999}, we take the following form for the viscous stress tensor $\op{\sigma}$:
\beq \sigma_{\alpha\beta} = \sigma_{\alpha\beta}\ut{yield}(\rho) + [\eta(\rho,\gdot) - \kappa(\rho,\gdot) \nabla^2] \left( \pd_\alpha u_\beta + \pd_\beta u_\alpha  - \frac{2}{d} \delta_{\alpha\beta} \pd_\gamma u_\gamma \right) + [\zeta(\rho,\gdot) - \kappa'(\rho,\gdot)\nabla^2] \delta_{\alpha\beta} \pd_\gamma u_\gamma,  
\label{eq_visc_stress_form}\eeq 
which corresponds to an expansion in gradients of the flow field $\uv$ respecting certain symmetry properties of the stress \footnote{The stress should remain invariant under coordinate inversion and change its sign whenever the spatial derivatives of the velocity field, i.e., the shear rate $\gdot$ and the compression rate $\nabla \cdot \uv$, change sign.}.
Here, $\eta$ and $\zeta$ denote the shear and bulk viscosity, respectively, which are generally functions of the density and the shear rate (see below).
The parameter $\kappa$ denotes the \emph{shear-curvature viscosity}, and the stress contribution associated with it serves to stabilize the flow field against large gradients. Analogously, the parameter $\kappa'$ controls the corresponding contribution stabilizing the bulk viscous stress.
The yield stress $\op{\sigma}\ut{yield}$ is independent of $\gdot$ and is nonzero only in the glassy phase ($\rho>\rho_g$). 
In contrast to the shear viscosity $\eta$ (see below), detailed data for the bulk viscosity $\zeta$ and the curvature viscosities $\kappa$ and $\kappa'$ in a hard-sphere fluid near the glass transition are not available.
Following Ref.\ \cite{jin_flow_2014}, we shall therefore assume these viscosities to have the same functional form as $\eta$, i.e.,
\begin{subequations}\begin{align}
\kappa(\rho,\gdot) &= \frac{\kappa_0}{\eta_0} \eta(\rho,\gdot), \\
\zeta(\rho,\gdot) &= \bvisc' \eta(\rho,\gdot), \\
\kappa'(\rho,\gdot) &= \bvisc' \kappa(\rho,\gdot).
\end{align}\label{eq_extra_visc}\end{subequations}
Here, $\eta_0$ and $\kappa_0$ are the shear (curvature) viscosities in the zero-shear rate (Newtonian) limit [see \cref{eq_visc_Newtonian}], and  $\bvisc'$ is a free dimensionless parameter. Typically, we set $\bvisc'=1$ and $\kappa_0/\eta_0\simeq (10-100)\psize^2$, where $\psize$ is a microscopic length scale, e.g., the average particle diameter in a colloidal glass. This choice gives rise to an effective interface width, $\sim \sqrt{\kappa_0/\eta_0}$, of the shear band of a few particle diameters $\psize$ \cite{jin_flow_2014}.
Using \cref{eq_shear_rt}, the relevant components of the viscous stress tensor follow as
\begin{align} \sigma_{xy} &= \sigma_{xy}\yield(\rho) + \eta(\rho,\gdot) \gdot - \kappa(\rho,\gdot) \pd_y^2 \gdot, \label{eq_sigma_xy}\\
\begin{split}\sigma_{yy} &= \sigma_{yy}\yield(\rho) + \left[ (\eta - \kappa \pd_y^2) \left(2-\frac{2}{d}\right) + (\zeta - \kappa'\pd_y^2)\right] \pd_y u_y, \\
&= \sigma_{yy}\yield(\rho) + \bvisc (\eta - \kappa \pd_y^2) \pd_y u_y,
\end{split}\label{eq_sigma_yy}\end{align}
with $\bvisc\equiv \bvisc'+4/3 = \sfrac{7}{3}$. 
In order to track the influence of the bulk viscosity, we shall carry along the parameter $\bvisc$ in our calculations.
Summarizing, \cref{eq_nse_red} reduces to
\begin{subequations}\begin{align}
\pd_t \rho &= -\pd_y (\rho u_y),\\
\pd_t (\rho u_x) &= \sigma_{\gdot} \pd_y \gdot + \sigma_\rho \pd_y\rho - \kappa \pd_y^3 \gdot - [(\pd_\rho \kappa) (\pd_y \rho) + (\pd_\gdot \kappa) (\pd_y \gdot)](\pd_y^2\gdot), \\
\pd_t (\rho u_y) &= -\Pi_{\gdot} \pd_y\gdot - \Pi_\rho \pd_y\rho + \bvisc(\eta - \kappa \pd_y^2) \pd_y^2 u_y,
\end{align}\label{eq_nse_simp2}\end{subequations}
where we defined
\begin{subequations}\begin{align}
\Pi_\rho &\equiv \pd (\Pi-\sigma_{yy}\yield) / \pd\rho,\qquad 
\Pi_{\gdot} \equiv \pd \Pi/\pd \gdot, \label{eq_Pi_deriv}\\
\sigma_\rho &\equiv \pd (\sigma\yield_{xy} + \eta \gdot) / \pd\rho,\qquad 
\sigma_{\gdot} \equiv \pd (\eta\gdot) / \pd \gdot, \label{eq_sigma_deriv}
\end{align}\end{subequations}
which are generally functions of $\rho$ and $\gdot$.

It seems reasonable to assume 
\beq \sigma\yield_{yy} \simeq \sigma\yield_{xy} = \sigma\yield,
\eeq
where $\sigma\yield$ is a common yield stress function.
In the liquid phase ($\rho< \rho_g$), the yield stress vanishes and the shear viscosity is well described by a Krieger-Dougherty relationship (cf.\ Ref.\ \cite{jin_flow_2014}):
\begin{subequations}\begin{align}
 \sigma\yield &= 0, \\
 \eta(\rho) &= \eta_0 (1-\Phi)^{-2}. \label{eq_visc_Newtonian}
\end{align}\end{subequations}
Here and in the following, $\Phi\equiv \rho/\rho_m$, where $\rho_m=0.67$ (in appropriate units, see below) is the packing fraction corresponding to random close packing of (polydisperse) hard spheres. 

In the glassy phase ($\rho>\rho_g$), instead, MD simulations of a hard-sphere system indicate \cite{mandal_heterogeneous_2012}  
\begin{subequations}\begin{align}
\sigma\yield(\rho) &= \frac{\sigma_0}{(1-\Phi)^p}, \\
\eta(\rho,\gdot) &= \sigma\yield(\rho) A(1-\Phi)^n \gdot^{n-1}, 
\end{align}\label{eq_constitut_glass}\end{subequations}
where the parameters  $\sigma_0 \simeq 0.0119\, \kbT/\psize^3$,  $A=34.5\,(\eta_0\psize^3/(\kbT))^{n}$, $p\simeq 2.355$, $n\simeq 0.4$ result from a fit.
$\kbT$ denotes the thermal energy and $\rho_g=0.585$ is the density of the glass transition.
The pressure is given, for any $\rho$, by \cite{mandal_heterogeneous_2012}
\beq 
\Pi(\rho,\gdot) = \frac{\Pi_0 \Phi}{(1-\Phi)}\left[ 1+ B(1-\Phi)^{1-r}\gdot^m\right],
\label{eq_constitut_press}\eeq 
with $\Pi_0\simeq 8.4\, \kbT/\psize^3$, $B=0.07\,(\eta_0\psize^3/(\kbT))^{m}$, $n=m\simeq 0.4$, $r=4.1$. The shear-rate dependence of $\Pi$ is a manifestation of the flow-induced distortion of the pair-correlation function.
We remark that the parameters in \cref{eq_constitut_glass,eq_constitut_press} have been obtained in Ref.\ \cite{mandal_heterogeneous_2012} from a fit to the global flow curves, taking $\gdot\equiv \gdot_0$, but are assumed here to apply also locally in the system.
We shall henceforth fix the units of mass, length and time by setting $\kbT=a=\eta_0=1$. 
With these choices, the fundamental ``microscopic'' time scale
$t_0 \equiv \eta_0 a^3/\kbT = 1$. Using the fact that $\eta_0$ is the fluid viscosity in the dilute limit [see \cref{eq_visc_Newtonian})] and invoking the Stokes-Einstein relation, one obtains $t_0 \sim a^2/D$ with the self-diffusion coefficient $D$. In other words, $t_0$ is the time needed for a particle to explore, in the dilute limit, a distance comparable to its own size. Noteworthy, this is also a measure of the structural relaxation time. Accordingly, the microscopic time scale $t_0$ determines, together with thermal energy and particle size, the viscosity and stress scale. In the context of macroscopic fluid dynamics, however, a more natural dimensionless measure of time, which we shall use in the discussion of our results, is instead given by the strain $t \gdot$.

\section{Linear stability analysis}

\subsection{Linearization of the dynamics}

We consider small fluctuations of the density and the shear-rate, i.e., $\rho(y,t) = \rho_0 + \delta\rho(y,t)$, $\gdot(y,t) = \gdot_0 + \delta \gdot(y,t)$, where  $\rho_0$ and $\gdot_0$ denote the uniform background values.
In linear order in the fluctuations and derivatives, \cref{eq_nse_simp2} becomes
\begin{subequations}\begin{align}
\pd_t \delta\rho &= -\rho_0 \pd_y u_y, \label{eq_lin_nse_rho}\\
\gdot_0\pd_t \delta\rho + \rho_0 \pd_t \delta\gdot  &= \sigma_{\gdot} \pd^2_y \delta \gdot + \sigma_\rho \pd^2_y \delta \rho - \kappa \pd_y^4 \delta \gdot, \label{eq_lin_nse_ux} \\
\rho_0 \pd_t u_y &= -\Pi_{\gdot} \pd_y \delta \gdot - \Pi_\rho \pd_y \delta \rho + \bvisc\eta \pd_y^2 u_y - \bvisc\kappa \pd_y^4 u_y, \label{eq_lin_nse_uy}
\end{align}\label{eq_lin_nse}\end{subequations}
where now the coefficients $\sigma_{\gdot,\rho}$, $\Pi_{\gdot,\rho}$, $\eta$, and $\kappa$ are understood to be evaluated for the background values $\rho_0$ and $\gdot_0$.

In order to develop a basic understanding of the transport mechanisms in the compressible fluid, note that, inserting \cref{eq_lin_nse_rho} into \cref{eq_lin_nse_ux}, the latter becomes a generalized diffusion equation for the shear rate fluctuation~$\delta\gdot$,
\beq \rho_0 \pd_t \delta\gdot  = \sigma_{\gdot} \pd^2_y \delta \gdot - \kappa \pd_y^4 \delta \gdot + \sigma_\rho \pd^2_y \delta \rho + \gdot_0 \rho_0 \pd_y u_y.
\label{eq_shear_rt_diffusion} \eeq 
While the last term on the r.h.s.\ is typically negligible, the first and the second term induce a smoothing of shear rate inhomogeneities. However, due to the third term, which is not present in a Newtonian fluid, a positive density fluctuation can effectively lower the local shear rate. Such a negative shear rate fluctuation drives, via the first term on the r.h.s. of \cref{eq_lin_nse_uy}, a flow which [via \cref{eq_lin_nse_rho}] further enhances the density in that region. This gives rise to a feedback mechanism, which is further analyzed in \cref{sec_lin_stab}.
In passing, we note that \cref{eq_lin_nse_rho,eq_lin_nse_uy} can be combined into a generalized ``sound-wave'' equation
\beq \pd_t^2 \delta\rho = \frac{\bvisc}{\rho_0} \left(\eta - \kappa\pd_y^2\right) \pd_y^2 \pd_t\delta\rho + \Pi_\gdot \pd_y^2\delta\gdot + \Pi_\rho\pd_y^2\delta\rho.
\label{eq_soundwave}\eeq 
The dynamics induced by the above compressible fluid equations is further discussed and contrasted to a diffusive transport model in \cref{app_transport}.

In order to investigate the linear stability, we solve \cref{eq_lin_nse} via the ansatz
\beq 
\bmat \delta\rho \\ \delta\gdot \\ u_y \emat
= 
\bmat \bar\rho \\ \bar \gdot \\ \bar u_y \emat
\exp(\omega t + \im k y),
\label{eq_fluct_ansatz}\eeq
where $\omega$ and $k$ represent the growth rate and wavenumber of a fluctuation, respectively, and the bared quantities denote the fluctuation amplitudes.
This ansatz transforms \cref{eq_lin_nse} into 
\begin{subequations}\begin{align}
\omega \bar \rho &= -\im  k \rho_0 \bar u_y, \label{eq_lin_nse1}\\
\omega (\gdot_0 \bar\rho + \rho_0 \bar \gdot) &= -k^2 (\sigma_{\gdot} + \kappa k^2) \bar \gdot -k^2 \sigma_\rho \bar \rho, \\
\omega \rho_0 \bar u_y &= -\im k \Pi_{\gdot} \bar \gdot - \im k \Pi_\rho \bar \rho - \bvisc( \eta + \kappa k^2) k^2 \bar u_y, 
\end{align}\label{eq_lin_nse_four}\end{subequations}
which can be written in matrix form as
\beq
\bmat 
\omega 			& 0 		& \im k \rho_0 \\
\omega \gdot_0 + k^2 \sigma_\rho 	& \omega \rho_0 + k^2 \tilde \sigma_{\gdot}(k)  & 0 \\
\im k \Pi_\rho 		& \im k \Pi_{\gdot}	& \omega \rho_0 + \theta(k) k^2
\emat  
\bmat \bar \rho \\ \bar \gdot \\ \bar u_y \emat 
= \bv{0},
\label{eq_linstab}
\eeq 
with the abbreviations
\beq \tilde \sigma_\gdot(k) \equiv \sigma_\gdot + \kappa k^2
\label{eq_sigma_k_def}\eeq
and
\beq \theta(k) \equiv \bvisc( \eta + \kappa k^2).
\eeq 
A nontrivial solution of eq.~\eqref{eq_linstab} requires the coefficient matrix to be singular and, correspondingly, the determinant to vanish:
\beq \rho_0 \omega^3 +  \left[\tilde\sigma_{\gdot}(k) + \theta(k) \right] k^2 \omega^2 + \left[\rho_0 \Pi_\rho +  \frac{\theta(k) k^2}{\rho_0}\tilde \sigma_\gdot(k) -\gdot_0 \Pi_\gdot \right] k^2 \omega + k^4 \left(\tilde \sigma_{\gdot}  \Pi_\rho - \sigma_\rho \Pi_{\gdot}\right) = 0.
\label{eq_detcond}
\eeq
Note that, in order to obtain a purely real solution, the ansatz in \cref{eq_fluct_ansatz} must be linearly combined with an expression of the same form but where $k$ is replaced by $-k$.
The three roots $w_{1,2,3}$ of the cubic equation \eqref{eq_detcond} are independent of the sign of $\pm k$.
Accordingly, we can write the general solution to the linearized hydrodynamic equations \eqref{eq_lin_nse} as
\beq \bmat \delta\rho \\ \delta\gdot \\ u_y \emat = \big(\bv{A} e^{w_1 t} + \bv{B} e^{w_2 t} + \bv{C} e^{w_3 t}\big)e^{\im k y} + \big(\bv{\hat A} e^{w_1 t} + \bv{\hat B} e^{w_2 t} + \bv{\hat C} e^{w_3 t}\big)e^{-\im k y}.
\label{eq_gensol}
\eeq
The coefficient vectors $\bv{A,B,}\ldots$ are obtained by inserting each root $\omega_j$ into \cref{eq_linstab} and determining the null-space of the resulting linear mapping.

\subsection{Boundary of stability and growth dynamics}
\label{sec_lin_stab}

Before turning to the discussion of the cubic equation \eqref{eq_detcond} in the full parameter space, we first focus on the region near the boundary of stability, where the analysis is simplified by the fact that the real part of at least one $\omega_j$ must be small.
We proceed by discussing the two possible cases admitted by the solutions to a cubic equation with real coefficients, like \cref{eq_detcond}.

\vspace*{2mm}
{\bfseries\underline{Case 1:}} All the three roots are purely real (but not necessarily distinct). The general solution given in \cref{eq_gensol} consists in this case only of exponentially growing, decaying or constant contributions. 
In the stable region, $\omega_j\leq 0$ for all $j$. Directly at the boundary to the unstable region, one must have $\omega_j=0$ for at least one mode index, say $j=1$.  Setting $\omega_1=0$ in \cref{eq_detcond} readily yields 
\beq \Bcal_1\equiv \tilde\sigma_{\gdot}(k) \Pi_\rho - \sigma_\rho \Pi_{\gdot}=0.
\label{eq_stab_bound1}
\eeq
As is shown below, the quantity $\Bcal_1$ defined here determines the boundary of stability. Inserting \cref{eq_stab_bound1} into \cref{eq_detcond}, the other two decay rates result as
\beq \omega_{2,3} = -\frac{\tilde \sigma_{\gdot}(k) + \theta(k)}{2\rho_0} k^2  \pm \sqrt{\left(\frac{\tilde \sigma_{\gdot}(k) + \theta(k)}{2\rho_0} k^2 \right)^2 - \left(\Pi_\rho+ \frac{\theta(k)k^2}{\rho_0^2}\tilde \sigma_\gdot-\frac{\gdot_0}{\rho_0}\Pi_\gdot \right) k^2}\,.
\label{eq_square_root} \eeq
For typical systems, one has 
\beq \rho_0 \Pi_\rho  \geq \gdot_0 \Pi_\gdot\,.
\label{eq_P_ineq_typ}\eeq 
In fact, for the constitutive relations reported in \cref{eq_constitut_glass,eq_constitut_press}, this inequality is violated only for unrealistically small shear rates $\gdot\lesssim 10^{-12}$ and extreme densities $\rho\simeq \rho_m$, where the hydrodynamic model considered here is doubtful.
Since generally $\tilde \sigma_{\gdot}\geq 0$ and $\theta(k)>0$,
it follows that $\omega_{2,3}\leq 0$ --- still assuming purely real $\omega_j$.
Accordingly, provided that \cref{eq_P_ineq_typ} holds, none of the frequencies $\omega_2$ and $\omega_3$ vanishes and, consequently, the boundary of stability is solely defined by the condition $\omega_1=0$ in this case.
Close to the boundary of stability, nonlinear terms in $\omega_1$ can be neglected in \cref{eq_detcond}, such that one readily obtains the growth rate
\beq \omega_1 \simeq - k^2 \frac{\Bcal_1}{\rho_0 \Pi_\rho + \theta(k) k^2 \tilde \sigma_\gdot(k) /\rho_0 - \gdot_0 \Pi_\gdot}. \;\;\;\;\;\;\;\text{(case 1, all frequencies real)}
\label{eq_growthrate_1}
\eeq
We thus infer that, under the condition in \cref{eq_P_ineq_typ}, the system is linearly stable if 
\beq \Bcal_1>0 \qquad \Leftrightarrow \qquad  \tilde \sigma_{\gdot}(k) \Pi_\rho >  \sigma_\rho \Pi_{\gdot}.
\label{eq_stab_cond1}\eeq
This inequality is consistent with the stability of the Navier-Stokes equations for a purely Newtonian fluid, since $\sigma_\rho=\Pi_{\gdot}=0$ and thus $\Bcal_1>0$ in that case.
As discussed below, \cref{eq_stab_cond1} in fact describes the boundary of stability of the whole relevant parameter space for the compressible single-component fluid.

In order for \cref{eq_gensol} to be real, one must have $\RE \bv{\hat A}= \RE\bv{A}$, $\IM \bv{\hat A} = -\IM \bv{A}$, with analogous conditions applying for $\bv{B}$ and $\bv{C}$.
These conditions are indeed fulfilled by the solution in \cref{eq_gensol}, which can be seen by writing \cref{eq_linstab} as 
\beq \left[\op{M}'(k) + \im \op{M}''(k)\right] \bv{A}=0,
\label{eq_linstab_rewrite}\eeq 
where $\op{M}'$ and $\op{M}''$ denote the real and imaginary parts of the matrix in \cref{eq_linstab}. 
Now let $\bv{A}=\bv{A'}+\im \bv{A''}$ be a solution to \cref{eq_linstab_rewrite}. 
Comparison of the real and imaginary parts of the resulting expression in \cref{eq_linstab_rewrite} shows that $\bv{\hat A}= \bv{A'} - \im \bv{A''}$ is a solution to the equation $\left[\op{M}'(-k) + \im \op{M}''(-k)\right] \bv{\hat A}= \left[\op{M}'(k) - \im \op{M}''(k)\right] \bv{\hat A}=0$, as required.

\vspace*{2mm}
{\bfseries\underline{Case 2:}} One root is real and the other two are complex conjugates.
Let $\omega_1$ denote the purely real and $\omega_{2,3} = \Omega'\pm \im \Omega''$ the complex conjugate solutions to eq.~\eqref{eq_detcond}.
For the imaginary part of \cref{eq_gensol} to vanish, $\bv{B}$ and $\bv{C}$ must be complex conjugates of one another, while $\bv{A}$ must be purely real. Taking $\bv{B} = \bv{C}^* = \bv{B'} +\im \bv{B''}$ allows one to write the general solution as
\beq (\rho , \gdot , u_y )^\mathrm{T} e^{-\im k y}  = \bv{A} e^{\omega_1 t} + 2 e^{\Omega' t} \left( \bv{B'} \cos \Omega'' t -  \bv{B''}\sin \Omega'' t \right).
\eeq
Analogously to case 1, at least either $\omega_1$ or $\Omega'$ must vanish at the boundary of stability.
If $\omega_1=0$, we recover eq.~\eqref{eq_stab_bound1} as a necessary consequence and eq.~\eqref{eq_square_root} shows that $\Omega'\leq 0$. Thus, in this case, the growing mode will be a monotonic function as in case 1 with a growth rate given by \cref{eq_growthrate_1}. In contrast, the oscillatory modes will in general be decaying functions of time and will not give rise to any linear instability.

In order to analyze the case $\Omega'=0$, we consider Vieta's formulas \cite{bronstein_handbook_2007} for the solutions to the cubic equation \eqref{eq_detcond}  in case 2:
\begin{subequations}
\begin{align}
 \omega_1+2\Omega' &=-k^2 \frac{\tilde \sigma_{\gdot} + \theta(k)}{\rho_0}, \label{eq_vieta1}\\
 2\omega_1 \Omega' + \Omega'^2 + \Omega''^2 &= k^2 \left(\Pi_\rho  + \frac{\theta(k)k^2}{\rho_0^2} \tilde\sigma_\gdot(k) -\frac{\gdot_0}{\rho_0}\Pi_\gdot \right), \label{eq_vieta2}\\
 \omega_1(\Omega'^2 + \Omega''^2) &= - \frac{k^4}{\rho_0}\left(\tilde \sigma_{\gdot} \Pi_\rho - \sigma_\rho \Pi_{\gdot}\right). \label{eq_vieta3}
\end{align}\label{eq_vieta}
\end{subequations}
If $\Omega'=0$, \cref{eq_vieta1} immediately implies $\omega_1<0$, i.e., the purely real mode is stable.
Moreover, combining the relations in \cref{eq_vieta1,eq_vieta2,eq_vieta3} results in 
\beq 
 \quad \omega_1 \Omega''^2 = -\frac{k^4}{\rho_0} ( \tilde \sigma_\gdot + \theta(k)) \left(\Pi_\rho + \frac{\theta(k)k^2}{\rho_0^2} \tilde\sigma_\gdot(k) -\frac{\gdot_0}{\rho_0}\Pi_\gdot\right) = -\frac{k^4}{\rho_0} \left(\tilde \sigma_{\gdot} \Pi_\rho - \sigma_\rho \Pi_{\gdot}\right) . 
\label{eq_vieta_cond}\eeq
In order to determine the stability boundary for the complex conjugate pair of solution, we consider in \cref{eq_detcond} small variations around $\Omega'=0$.
Accordingly, we insert $\omega = \delta\Omega' \pm \im \Omega''$ into  \cref{eq_detcond}, where $\Omega''$ is determined by \cref{eq_vieta2}.  
Neglecting terms of $\Ocal(\delta\Omega'^2)$ and higher in \cref{eq_detcond} (keeping, however, all orders in $\Omega''$, as this quantity is not necessarily small), yields
\beq \delta\Omega'  \simeq - \frac{1}{2} \frac{\rho_0 k^2 \Bcal_2}{\left\{ \Pi_\rho \rho_0^2 + k^2 [\theta(k)^2 + 3\theta(k)\tilde\sigma_\gdot(k) + \tilde\sigma_\gdot(k)^2] - \rho_0\gdot_0 \Pi_\gdot\right\}},
\label{eq_growthrate_2}\eeq
with
\beq\begin{split} \Bcal_2 &\equiv \left[ \tilde \sigma_\gdot + \theta(k)\right] \left(\Pi_\rho + \frac{\theta(k)k^2}{\rho_0^2} \tilde\sigma_\gdot(k) -\frac{\gdot_0}{\rho_0}\Pi_\gdot\right) -  \left(\tilde \sigma_{\gdot} \Pi_\rho - \sigma_\rho \Pi_{\gdot}\right), \\
&= \theta(k) \left(\Pi_\rho + \frac{\theta(k)k^2}{\rho_0^2} \tilde\sigma_\gdot(k) -\frac{\gdot_0}{\rho_0}\Pi_\gdot\right) + \tilde\sigma_\gdot \left( \frac{\theta(k)k^2}{\rho_0^2} \tilde\sigma_\gdot(k) -\frac{\gdot_0}{\rho_0}\Pi_\gdot\right) + \sigma_\rho \Pi_\gdot. 
\end{split}
\label{eq_stab_bound2}\eeq 
Under the condition \eqref{eq_P_ineq_typ}, the denominator on the r.h.s.\ in \cref{eq_growthrate_2} is positive for all $k$, allowing one to conclude that the system is linearly stable for 
\beq \Bcal_2>0.
\label{eq_stab_cond2}\eeq
As expected, the condition $\Bcal_2=0$ coincides with \cref{eq_vieta_cond}.
Note furthermore that $\Bcal_2>0$ for $k\to\infty$ and, generally, $\Bcal_2 > \Bcal_2|_{k=0} = \theta(0) \left(\Pi_\rho - \gdot_0 \Pi_\gdot/\rho_0\right) + \Pi_\gdot \left(\sigma_\rho  - \gdot_0 \sigma_\gdot/\rho_0 \right)$.
A numerical analysis reveals that the condition $\Bcal_2|_{k=0}>0$ and thus $\Bcal_2(k)>0$ is fulfilled for all physically relevant $\rho_0$ and $\gdot_0$ of the present model.

The main result of the above analysis is that the cubic equation \cref{eq_detcond} admits  instability only through a single monotonically growing mode, the two other modes being decaying functions of time, either in a monotonic (case 1) or an oscillatory (case 2) fashion.

\subsection{Stability diagram and discussion}
\label{sec_stabdiag}

\begin{figure}[t]\centering
    \subfigure[]{\includegraphics[width=0.5\linewidth]{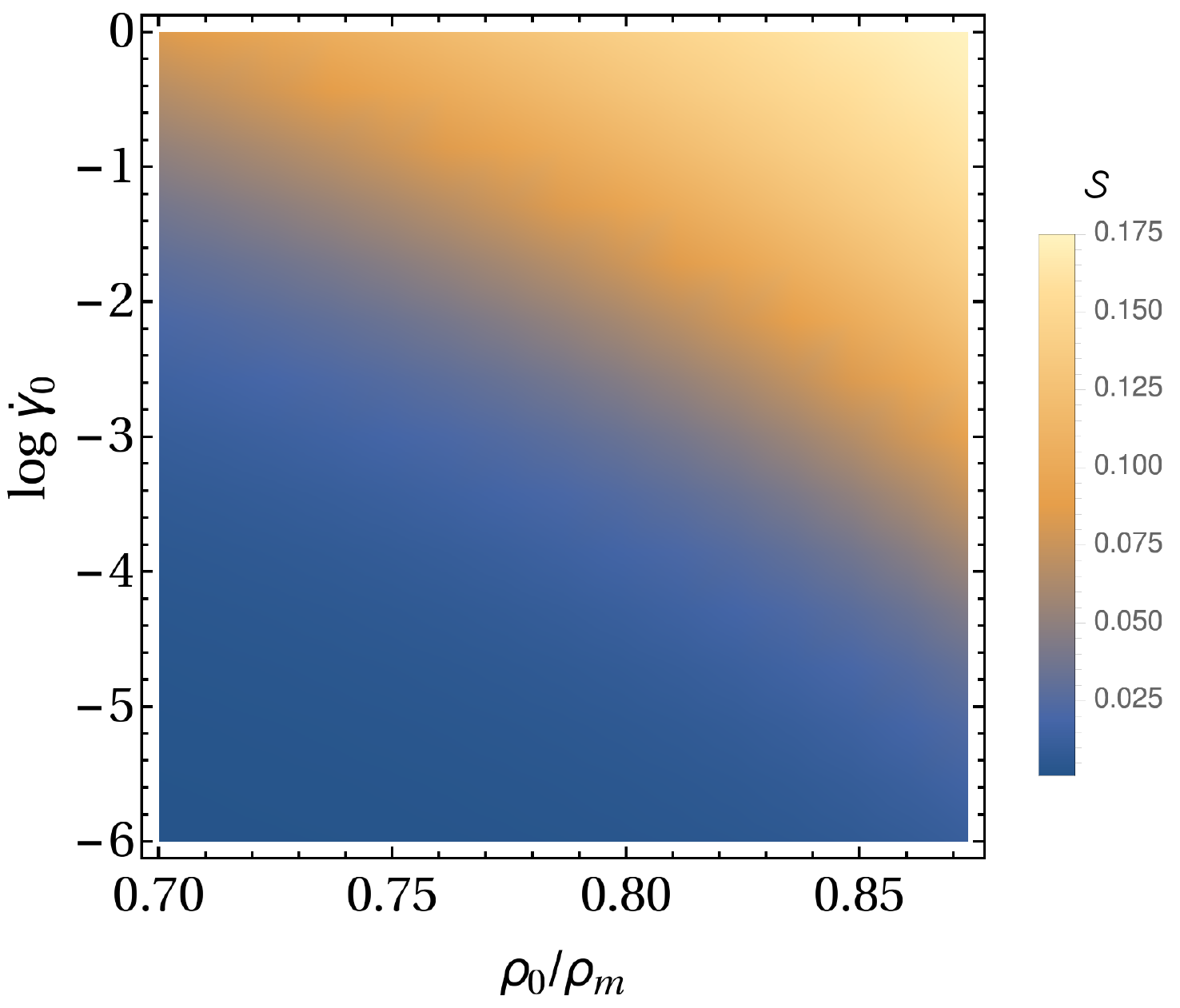} \label{fig_stabdiag_liq}}\hfill
    \subfigure[]{\includegraphics[width=0.48\linewidth]{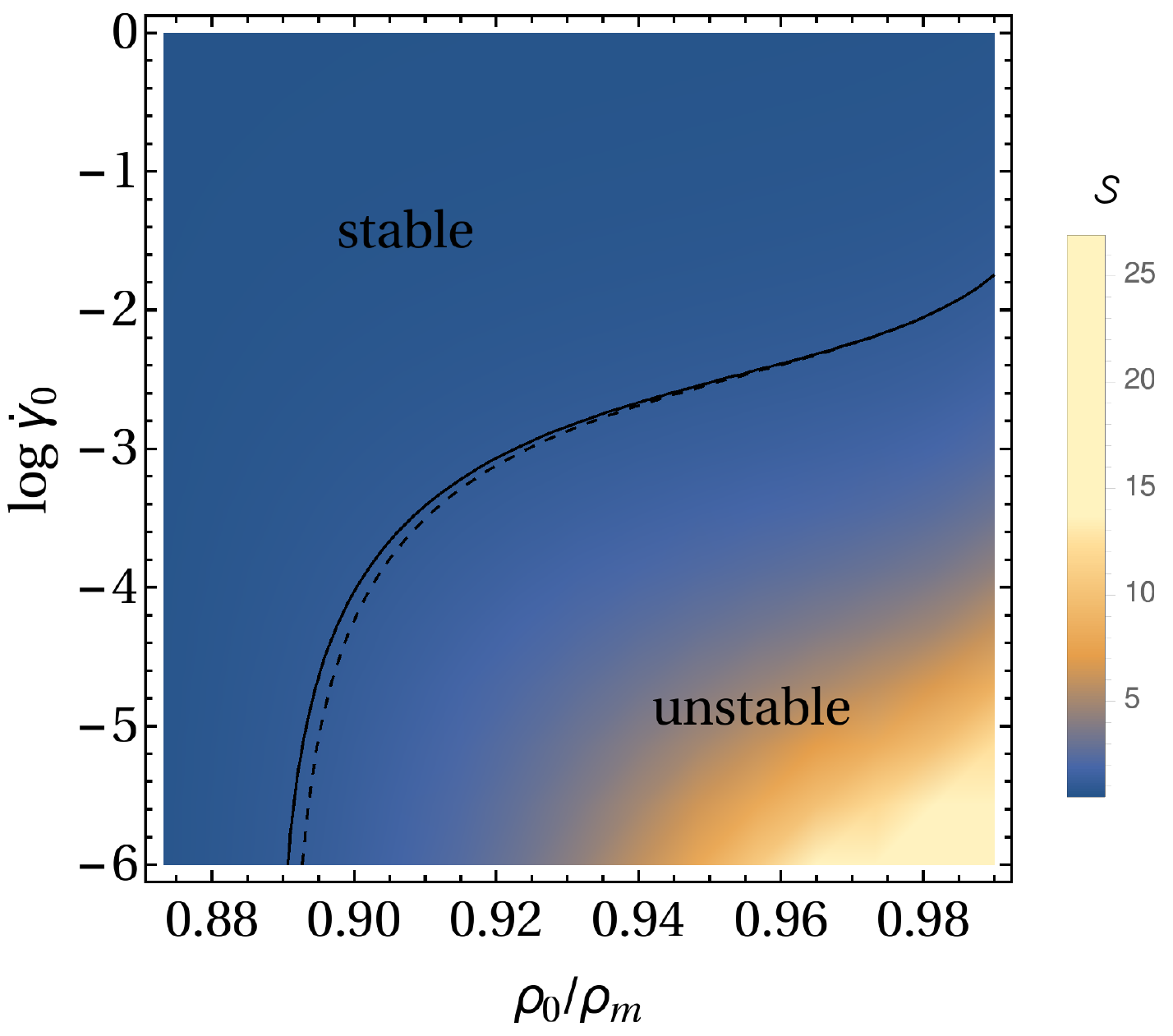} \label{fig_stabdiag_glass}}
    \caption{Stability diagram for a liquid (a) below ($\rho<\rho_g$) and (b) above ($\rho>\rho_g$) the glass transition. The glass transition occurs at a density of $\rho_g/\rho_m \simeq 0.873$, where $\rho_m=0.67$ denotes the density of random close packing (in dimensionless units, see \cref{sec_model}). The SDC instability occurs for values of the parameter $\Scal>1$ [\cref{eq_stab_param}] or, equivalently, for $\Bcal_1<0$ [\cref{eq_stab_cond1}]. The boundary of stability is indicated by the solid curve in (b), corresponding to $\Scal=1$. For comparison, the dashed curve in (b) represents the boundary of stability computed with $\sigma_{yy}\yield=0$ in $\Pi_\rho$ [\cref{eq_Pi_deriv}].}
    \label{fig_stabdiag}
\end{figure}

\begin{figure}[t]\centering
    \subfigure[]{\includegraphics[width=0.45\linewidth]{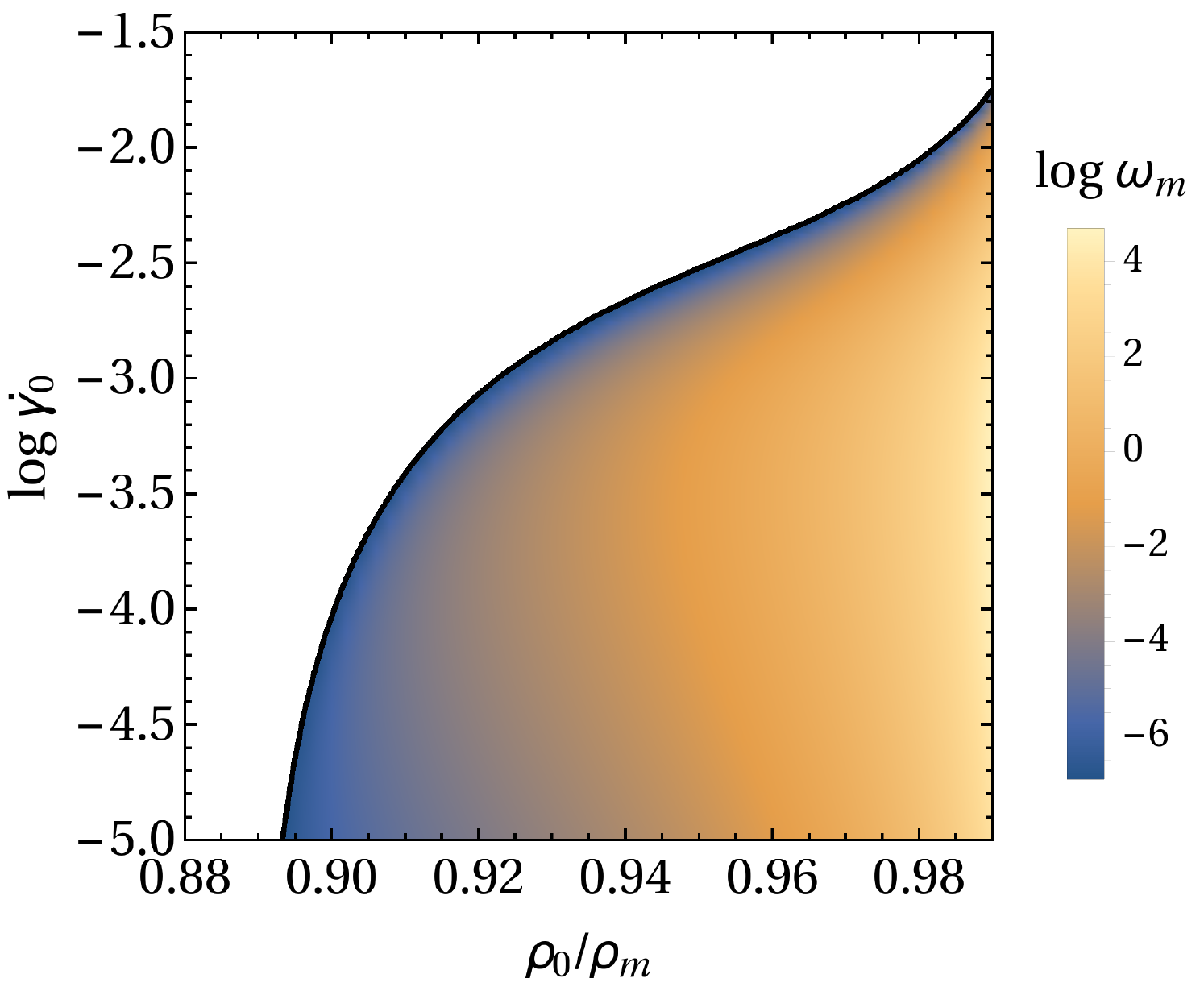} \label{fig_wmax}}\hfill
    \subfigure[]{\includegraphics[width=0.45\linewidth]{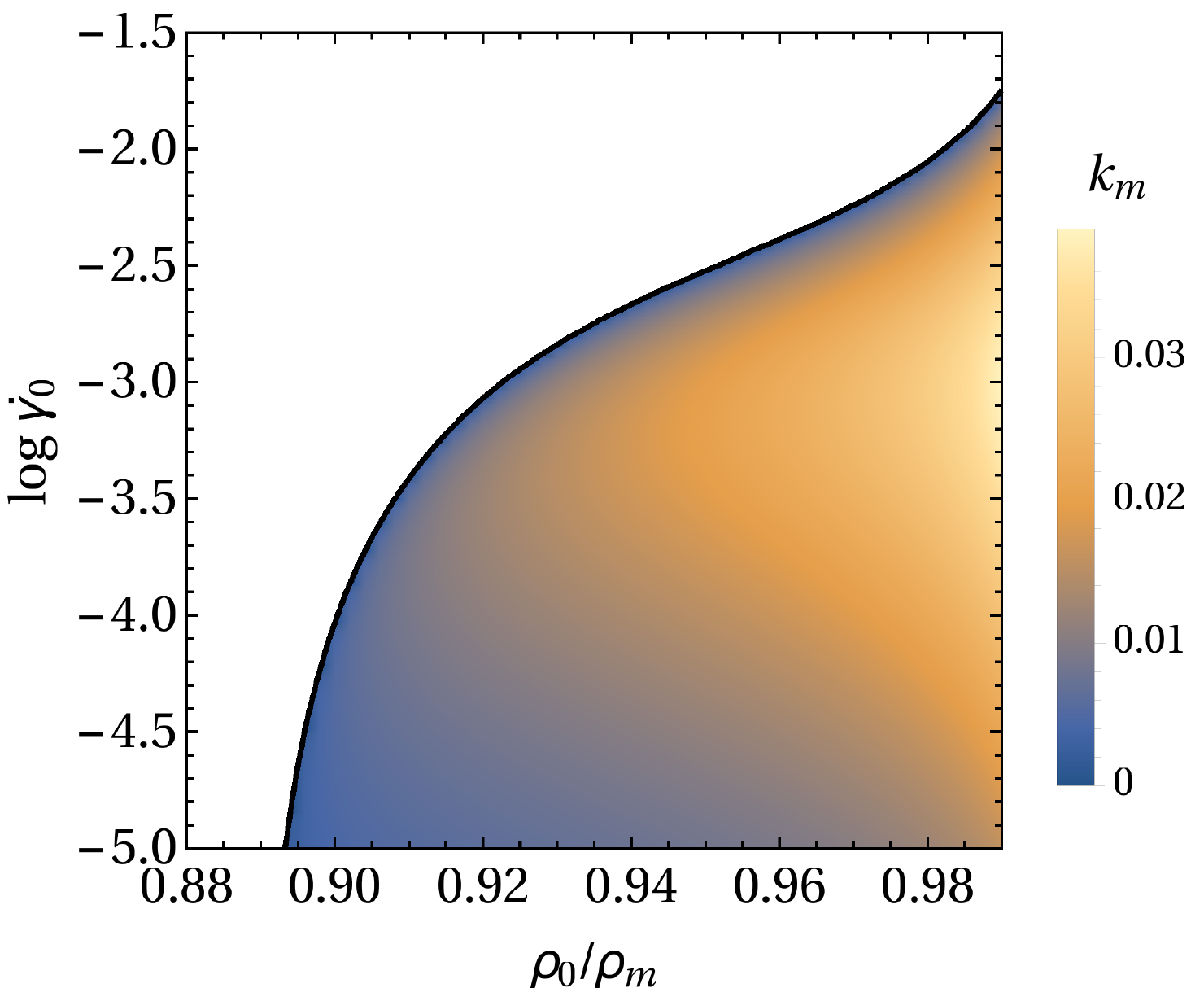} \label{fig_kmax}}  
    \subfigure[]{\includegraphics[width=0.435\linewidth]{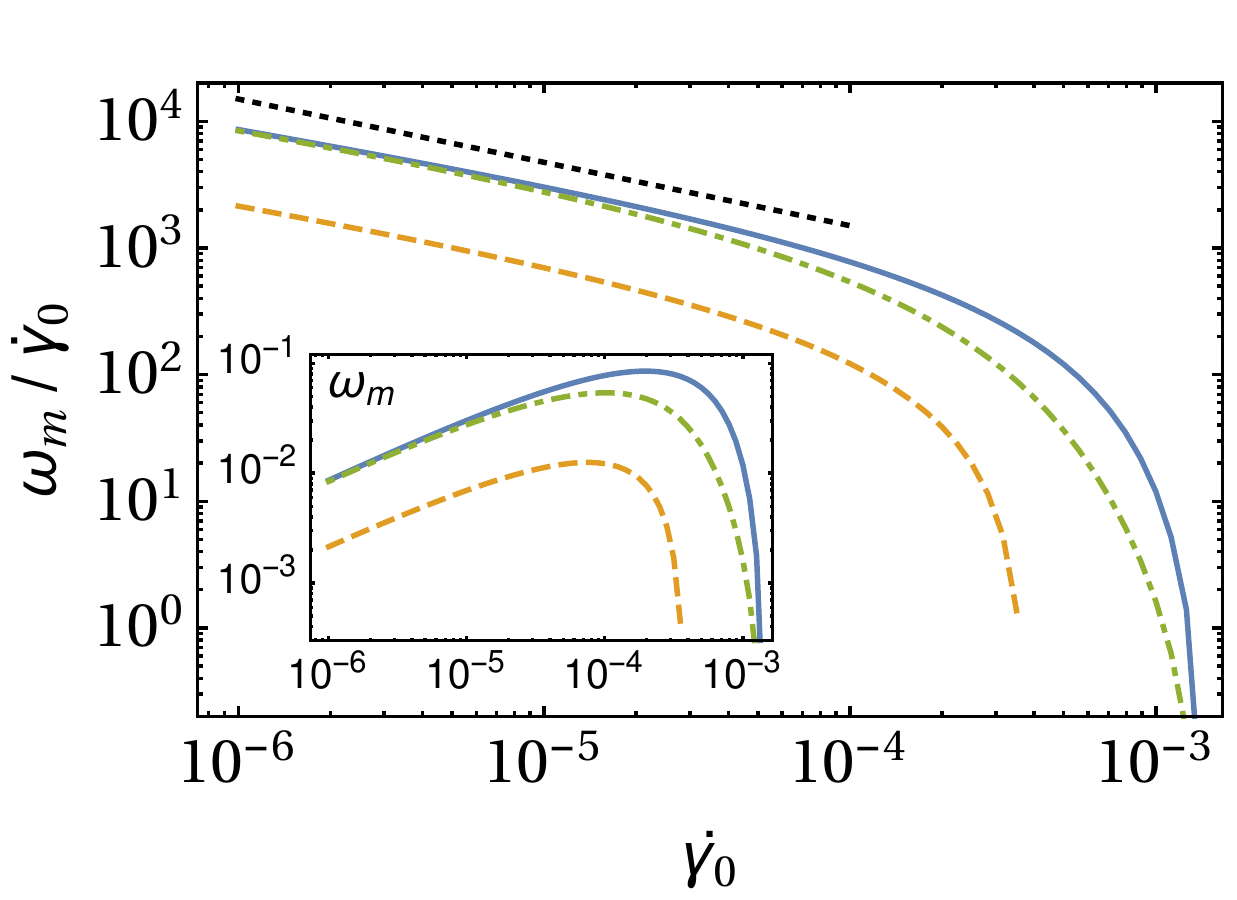}     \label{fig_growthrate}}\hfill
    \subfigure[]{\includegraphics[width=0.465\linewidth]{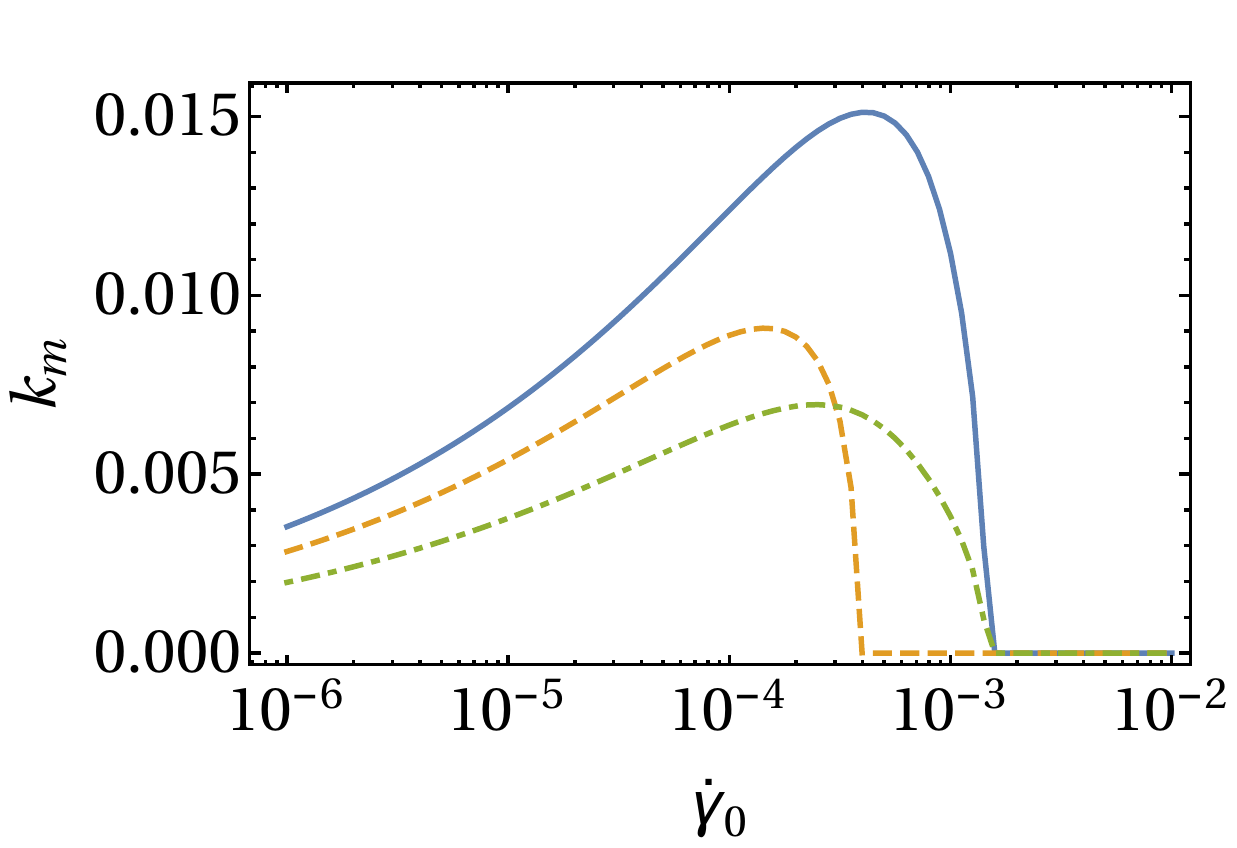}     \label{fig_growthmode}}
    \caption{(a,b) Maximum growth rate $\omega_m$ (a) and associated wavenumber $k_m$ (b) of the unstable modes. In the white region, the system is stable ($\omega<0$). A value of $\kappa_0=100$ is used for the calculation. (c,d) Growth rate $\omega_m$ (c) and wavenumber $k_m$ (d) of the maximally unstable mode as a function of $\gdot$ for $\rho_0/\rho_m=0.93$, $\kappa_0=100$ (solid curve), $\rho_0/\rho_m=0.91$, $\kappa_0=100$ (dashed curve), and $\rho_0/\rho_m=0.93$, $\kappa_0=1000$ (dot-dashed curve). The dotted line in (c) represents a power-law $\propto \gdot^{-0.5}$. The main plot and the inset in (c) shows $\omega_m$ expressed in terms of the inverse shear rate $1/\gdot_0$ and the microscopic time scale $t_0$, respectively (see \cref{sec_model}).}
    \label{fig_rate}
\end{figure}

\begin{figure}[t]\centering
    \includegraphics[width=0.45\linewidth]{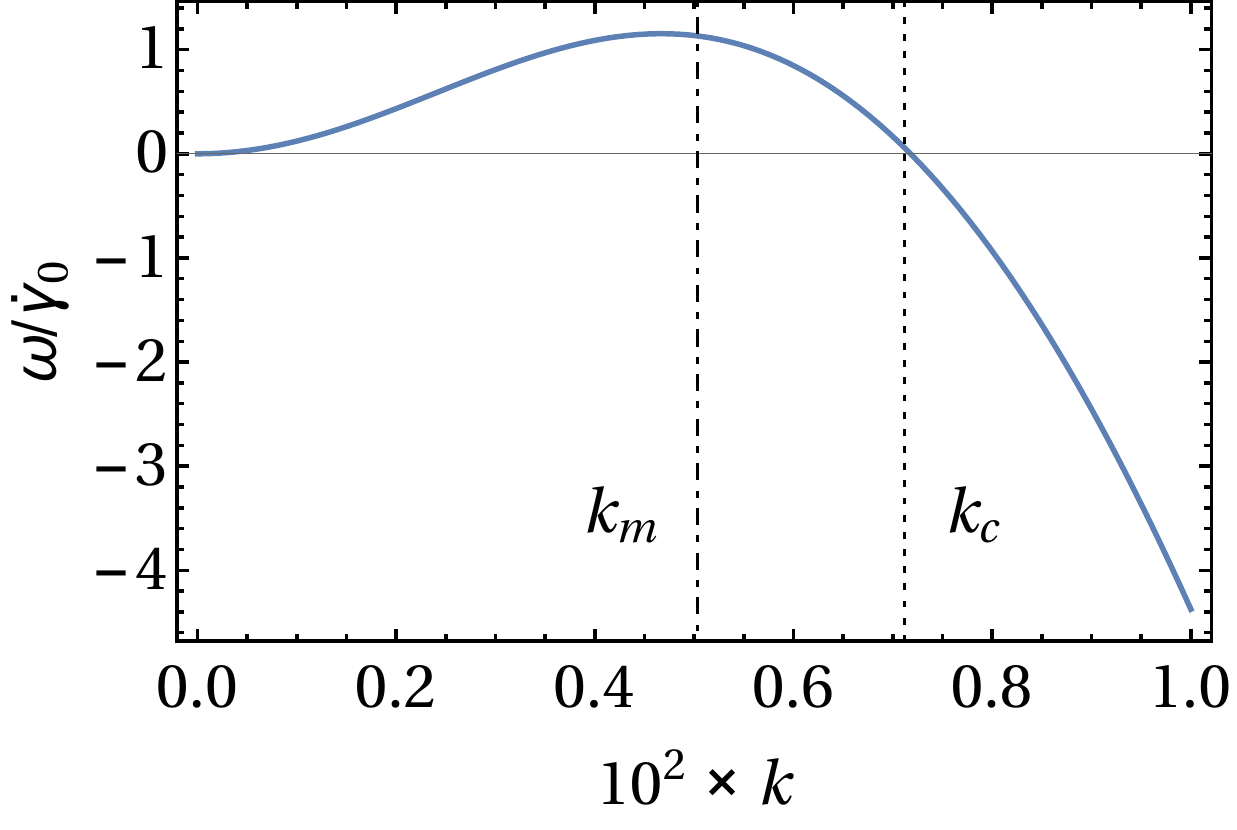}  
    \caption{(c) Typical behavior of the growth rate $\omega$ [\cref{eq_detcond}] as a function of the wavenumber $k$ in the unstable region of the parameter space. For wavenumbers $k$ with $0<k<k_c$, the system is unstable. The dotted line indicates the location of the critical wavenumber $k_c$ [\cref{eq_k_crit}]. Near the boundary of stability, the wavenumber $k_m$ of the fastest growth mode is estimated by \cref{eq_k_max} (dash-dotted line). Near the boundary of stability and for small $k$, the growth rate $\omega$ is well approximated by \cref{eq_growthrate_1}, implying $\omega\propto k^2$. The values $\rho_0=0.91\rho_m$, $\gdot_0\simeq 3.5\times 10^{-4}$, and $\kappa_0=100$ are used for the calculation.}
    \label{fig_disprel}
\end{figure}

As shown above, within the present linear stability analysis of the hydrodynamic equations for a compressible single component yield-stress fluid, the instability occurs uniquely via a monotonic (and thus non-oscillatory) mode, which grows exponentially with a rate given by \cref{eq_growthrate_1}.
In fact, an extensive numerical evaluation of the solutions of the dispersion relation in \cref{eq_detcond} indicates that, over the whole relevant parameter space, \emph{all} unstable modes have a non-oscillatory character. Owing to \cref{eq_P_ineq_typ} and the fact that $\Bcal_1\sim k^2>0$ in the limit $k\to \infty$, the growth rate in \cref{eq_growthrate_1} becomes negative for sufficiently large $k$, as is necessary for a physically reasonable model. 
In particular, asymptotically for $k\to\infty$ one obtains
\beq \omega(k\to\infty)\simeq - \frac{\rho_0 \Pi_\rho}{\bvisc \kappa k^2}.
\label{eq_rate_large_k}\eeq 
Note, however, that the continuum model in \cref{eq_nse} is not expected to be valid at arbitrarily small scales.
We remark that, in the absence of the shear- and bulk-curvature viscosities $\kappa$ and $\kappa'$, one has $\omega(k\to\infty) \simeq -\rho_0 \Bcal_1/(\bvisc\eta \sigma_\gdot)$.
On the other hand, neglecting all contributions related to bulk viscosity, yields $\omega(k\to\infty)\simeq - k^4 \kappa \Pi_\rho /(\rho_0 \Pi_\rho - \gdot_0 \Pi_\gdot) - k^2 \sfrac{\Bcal_1}{(\rho_0 \Pi\rho - \gdot_0 \Pi_\gdot)} $.
From these results one infers that the stability of the system for large $k$ is indeed due to the shear-curvature viscosity.

Since $\tilde \sigma_\gdot$ is a growing function of $k$, the \emph{boundary of stability} of the whole phase diagram of a bulk system is determined by \cref{eq_stab_cond1} for $k=0$. Specifically, if $\Bcal_1(k=0)<0$, the system is unstable for all wavenumbers, satisfying $\Bcal_1(k)=\Bcal_1(k=0)+\kappa  \Pi_\rho k^2 <0$ (cf.\ \cref{eq_growthrate_1}). In other words, all wavenumbers $0< k < k_c$ are unstable, where the critical wavenumber $k_c$ is defined via 
\beq 
k_c = \sqrt{-\frac{\Bcal_1(k=0)}{\kappa\Pi_\rho}}.
\label{eq_k_crit}
\eeq  
Remarkably, this expression as well as the condition for stability threshold, $\Bcal_1(k)=0$, are identical to the corresponding expressions obtained in Ref.\ \cite{jin_flow_2014} for the advection-diffusion model described by \cref{eq_advdiff,eq_Stokes} \footnote{The actual shape of the boundary of stability is different in Ref.\ \cite{jin_flow_2014} owing to the use of different constitutive equations.}.
However, as can be inferred from \cref{eq_growthrate_1}, owing to the presence of bulk viscosity, the fastest growing mode behaves differently as a function of $k$ for the compressible fluid. Such a finite bulk viscosity is to be expected, since colloidal suspensions exhibit a certain degree of local compressibility even in the highly concentrated regime.
\Cref{fig_stabdiag} shows the stability diagram obtained for $k=0$. 
For illustrative purposes, it is more convenient to consider instead of $\Bcal_1$ [\cref{eq_stab_bound1}] the (dimensionless) stability parameter
\beq \Scal \equiv  \frac{\sigma_\rho \Pi_\gdot }{\sigma_\gdot \Pi_\rho},
\label{eq_stab_param}\eeq  
according to which the system is unstable for values $\Scal>1$.
As seen in \cref{fig_stabdiag}, the instability occurs only in the glassy phase ($\rho>\rho_g$).

The wavenumber $k_m$ and the growth rate $\omega_m$ of the fastest growing mode has to be determined numerically from \cref{eq_detcond} in the general case.
\Cref{fig_rate} shows $k_m$ and $\omega_m$ as functions of the background density $\rho_0$ and shear rate $\gdot_0$. 
When expressed in terms of the fundamental time and length scales $t_0$ and $a$ (which are unity for our choice of units), $\omega_m$ and $k_m$ reach a maximum for intermediate shear rates and generally grow upon increasing the density.
When taking instead the inverse shear rate as the fundamental time scale, $\omega_m/\gdot_0$ grows with increasing distance from the boundary of stability [see main plot of \cref{fig_growthrate}]. At intermediate shear rates, an effective algebraic behavior $\omega_m/\gdot_0 \sim \gdot_0^{-0.5}$ can be inferred from the numerics. 
As illustrated in \cref{fig_growthrate,fig_growthmode}, changing the value of the shear-curvature parameter $\kappa_0$ [\cref{eq_extra_visc}] has only a moderate effect on $k_m$ and $\omega_m$.

Close to the stability boundary, $\Bcal_1$ [\cref{eq_stab_bound1}] and therefore $k_c$ are small, such that a Taylor expansion of the growth rate in \cref{eq_growthrate_1} to $\Ocal(k^4)$ is sufficient to determine $k_m$. (The leading term of the expansion of $\omega$ is of $\Ocal(k^2)$.) Within this approximation, the wavenumber of the fastest growing mode follows by evaluating the condition $\d\omega/\d k =0$ as 
\beq k_m \simeq \sqrt{\frac{-\Bcal_1^{(0)} \rho_0 (\rho_0 \Pi_\rho - \gdot_0 \Pi_\gdot)}{ 2 \kappa \rho_0 \Pi_\rho (\rho_0 \Pi_\rho - \gdot_0 \Pi_\gdot) -2 \bvisc \eta \sigma_\gdot \Bcal_1^{(0)} }} \simeq \sqrt{\frac{-\Bcal_1^{(0)} }{2\kappa \Pi_\rho}},
\label{eq_k_max}
\eeq 
with $\Bcal_1^{(0)}= \Bcal_1|_{k=0}$.
In the last expression, the fact that $\Bcal_1^{(0)}\simeq 0$ close to the stability boundary has been used.
Notably, due to bulk viscous effects, \cref{eq_k_max} is generally different from the corresponding result obtained in Ref.\ \cite{jin_flow_2014}.
In \cref{fig_disprel}, the typical behavior of the growth rate $\omega$ as a function of the wavenumber $k$ is illustrated. We have chosen here values of the parameters $\rho_0$ and $\gdot_0$ near the stability boundary, where \cref{eq_k_max} provides an accurate approximation to the actual maximum wavenumber.
While $\omega\propto k^2$ for small $k$ [see \cref{eq_growthrate_1}], $\omega$ eventually becomes negative for sufficiently large $k$ [see \cref{eq_rate_large_k}], as required for reasons of stability.

\begin{figure}[t]\centering
    \includegraphics[width=0.45\linewidth]{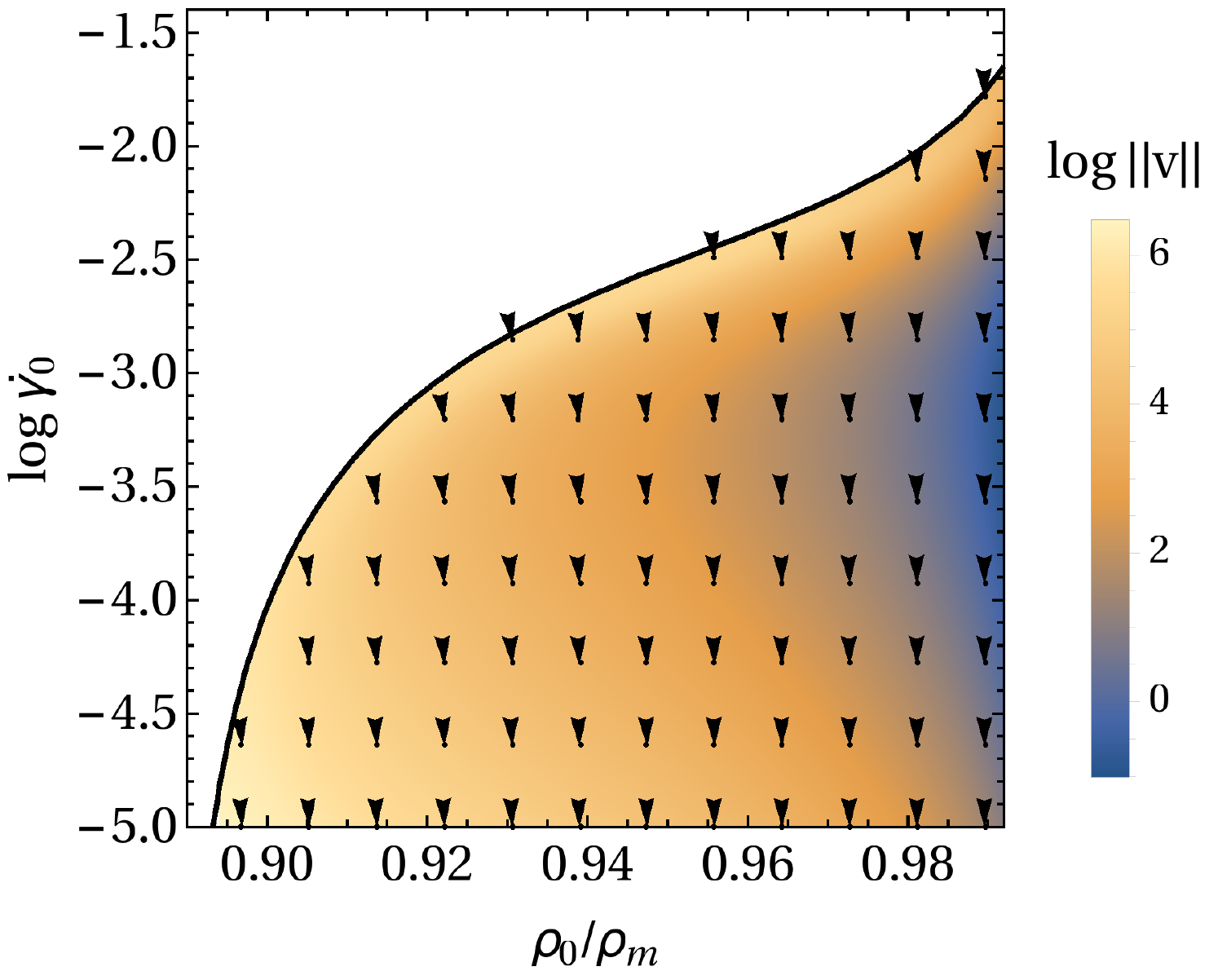}  
    \caption{Growth direction and magnitude of the fluctuation amplitude $\vv \propto (\bar\rho/\rho_0, \bar{\gdot}/\gdot_0)$ (up to a normalization factor, see text) of the most unstable mode $k_m$, as determined by the nullspace solution of \cref{eq_fluct_ansatz}. The coloring indicates the magnitude of $\vv$ in a logarithmic scale, while the arrows indicate the growth direction (non-logarithmic scale along both axes), which is determined up to a sign. Accordingly, in the unstable region, fluctuations grow indefinitely by reducing the shear rate and increasing the density (or vice versa), providing a nonlinear feedback mechanism for the SDC-instability.}
    \label{fig_fluctdir}
\end{figure}

\Cref{fig_fluctdir} illustrates the direction of growth and the magnitude of the most unstable mode (having wavenumber $k_m$). In order to obtain the amplitude vector $\vv(\rho_0,\gdot_0)$, the nullspace solution $\vv_0 = (\bar\rho, \bar{\gdot}, \bar u_y)$ of \cref{eq_linstab} is determined and normalized, $\vv_0/||\vv_0||$, and then projected onto the space spanned by $\rho_0$ and $\gdot_0$, additionally normalizing the components by $\rho_0$ and $\gdot_0$, respectively. As illustrated in \cref{fig_fluctdir}, in general, $\bar\rho$ and $\bar{\gdot}$ have opposite signs in the unstable region, as expected for the SDC instability. A similar anti-correlation has been reported in molecular dynamics studies of heterogeneous flow in a hard sphere glass~\cite{mandal_heterogeneous_2012}. 
Note that, with $\vv$, also $-\vv$ is a valid solution of \cref{eq_linstab}; in the plot, we have chosen the positive sign of $\bar\rho$. One notes that the development of the instability is dominated by a strong relative change of the shear rate, while the growth of the density is rather weak.
This feature of the linear regime will also prevail in the nonlinear case discussed below.
Comparing with \cref{fig_wmax,fig_kmax}, one infers that the growth amplitude $||\vv||$ is largest in those those regions of the phase diagram where the growth rate $\omega_m$ and the wavenumber $k_m$ are relatively small.

\section{Nonlinear dynamics and steady states}
\label{sec_nonlin}

\subsection{Dynamics}

\begin{figure}[t]\centering
	\subfigure[]{\includegraphics[width=0.33\linewidth]{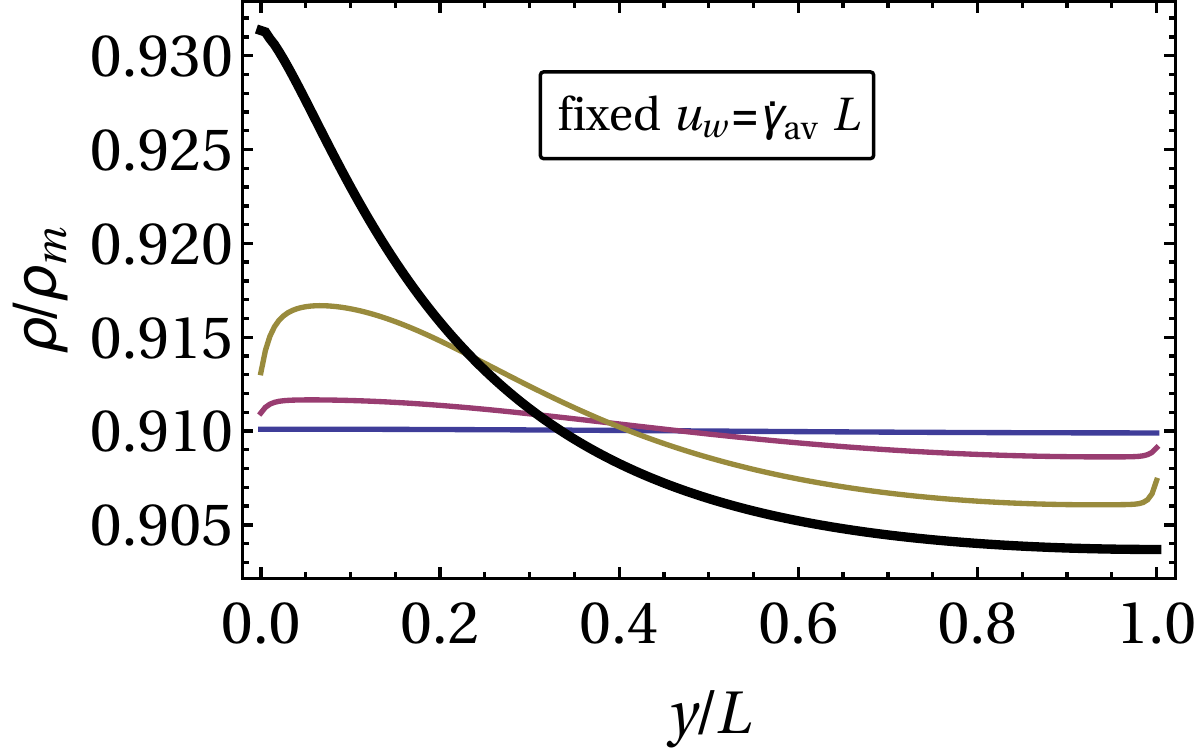} }\hfill
	\subfigure[]{\includegraphics[width=0.32\linewidth]{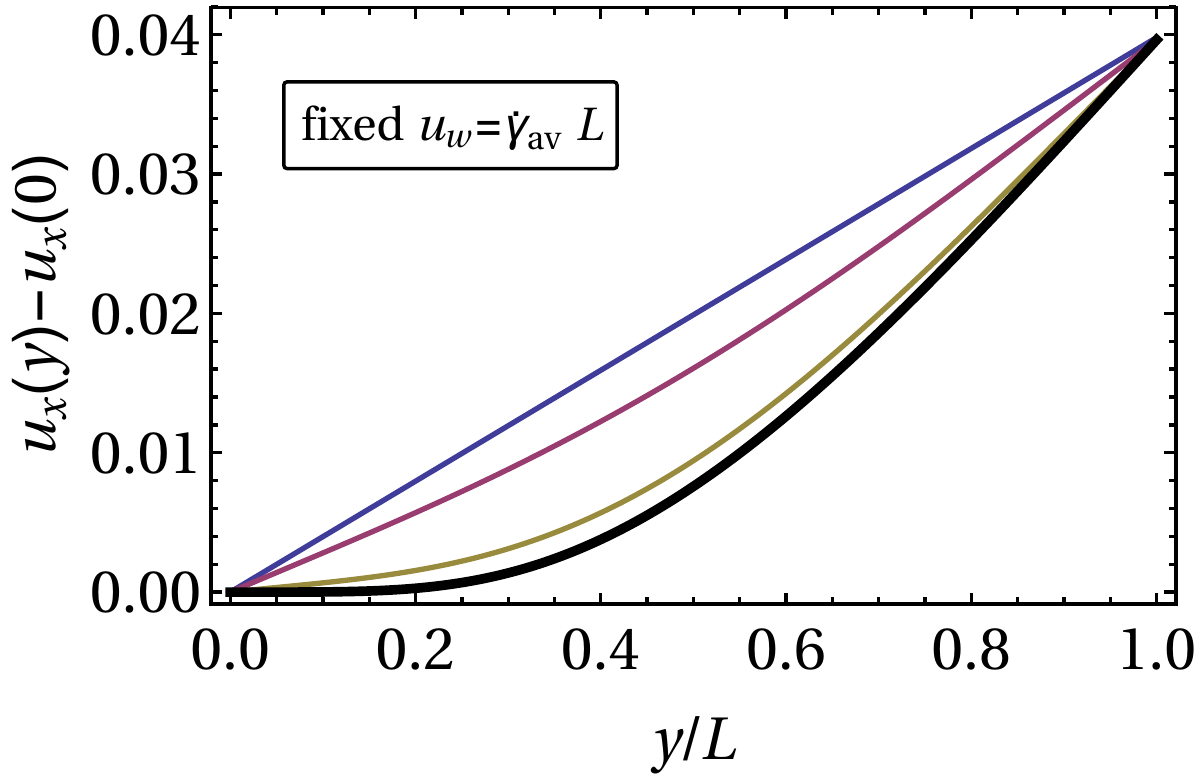} } \hfill
	\subfigure[]{\includegraphics[width=0.31\linewidth]{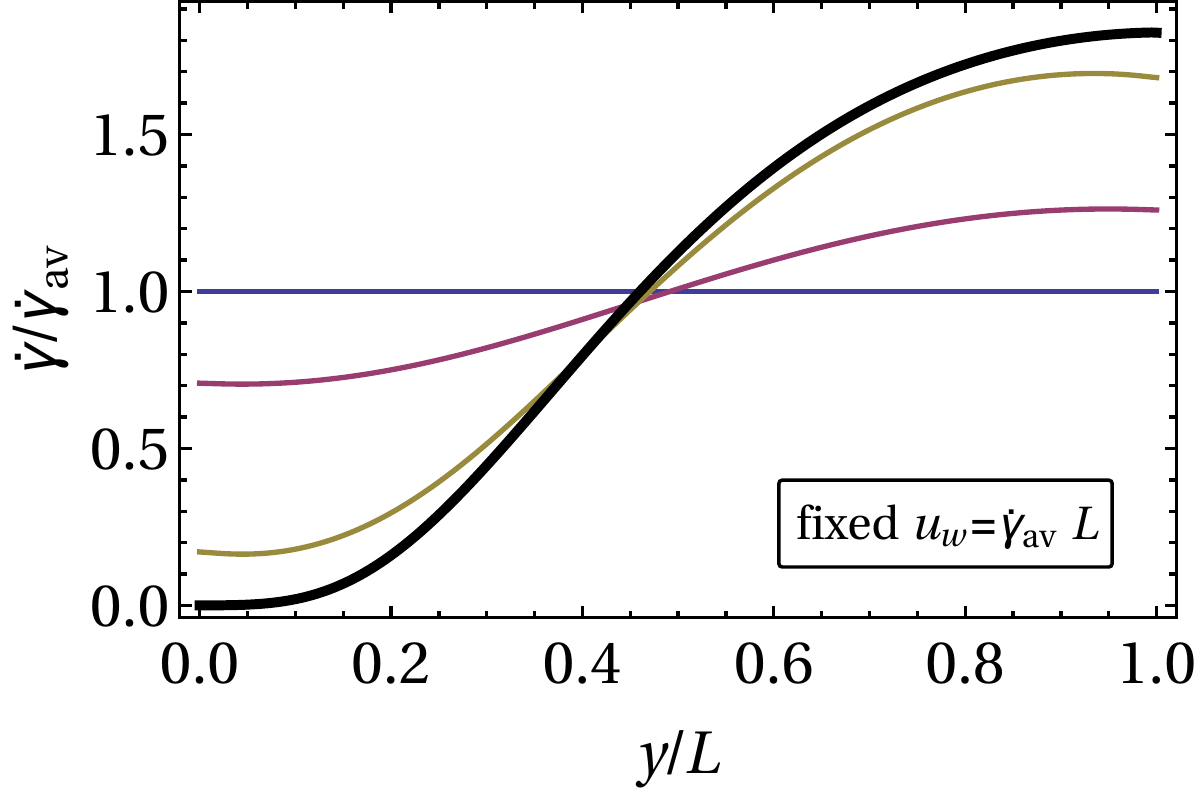} } 
	\caption{Time evolution of (a) the density, (b) the velocity, and (c) the shear rate resulting from \cref{eq_nse_red} for a fixed wall velocity $u_w=u_x(L)-u_x(0)$ corresponding to an average shear rate $\gdot\st{av}=2\times 10^{-4}$. The profiles shown are obtained at times $t\gdot\st{av} =0, 0.20, 0.28,\infty$, where $t=\infty$ corresponds to the steady state reached for $t\gdot\st{av} \gtrsim 10^3$ (thick black curve). In the steady state, the local shear rate at the left and the right boundary are found to be $\gamma\simeq 1.7\times 10^{-8}$ and $3.7\times 10^{-4}$, respectively. The initial growth rate of the maximally unstable mode is given by $\omega_m\simeq 7.8\times 10^{-3}$ [see \cref{eq_detcond}]. Parameters $L=200\Delta h$, $\kappa_0=100$ and $\rho_0=0.91\rho_m$ are used.}
	\label{fig_dyn_fixedU}
\end{figure}

The one-dimensional Navier-Stokes equations for a compressible fluid given in \cref{eq_nse_red} are numerically solved in a slit geometry (see \cref{fig_setup}) in the following way: the flux $\bv{j}=\rho\uv$ is introduced and the partial differential equations are converted to ordinary ones by spatial discretization on a grid of $L/\Delta h$ nodes \cite{moin_fundamentals_2010,mazumder_numerical_2015}. Specifically, we use second-order accurate central differences for the approximation of the spatial derivatives. The grid spacing is taken as $\Delta h=a$, which is thus unity in our choice of units. 
A vanishing normal flux $j_y$ is assumed at the boundaries.
The lateral flux $j_x$ at the boundaries is determined by imposing, at both walls, either a constant wall velocity $u_w$, a constant wall shear rate $\gamma_w$, or a constant wall stress $\sigma_w$.
Values of $j_x$ exterior to the computational domain (``ghost nodes'') are calculated via linear interpolation from the adjacent bulk nodes \footnote{The resulting steady state profiles depend sensitively on the boundary conditions. If, in the case of a fixed wall velocity, one, e.g., imposes a constant velocity in the exterior nodes (implying a vanishing wall shear rate), a symmetric steady state shear-rate profile is found, cf.\ \cref{fig_dyn_fixedU}.}.
Exterior values of $\rho$ are determined by assuming a vanishing gradient of $\rho$ at the boundary.
We have checked that the total density, $\int_0^L \d y\, \rho(y)$, remains practically constant during the time evolution.

As initial configuration we use a density and shear rate profile with a weak sinusoidal modulation (barely visible in the plots) in order to trigger the SDC instability.
In the case of fixed wall stress, we initialize the shear rate with the constant value $\gdot_0$ calculated from \cref{eq_steady_shearDensity_rel} below. 
The dynamical evolution is, however, not significantly altered if instead a different value for the initial shear rate is used, except for a short transient at early times.
After this transient, the evolution of the shear rate is found to be essentially enslaved to the density dynamics.
In all cases, the wavelength of the maximally unstable mode predicted by the dispersion relation [\cref{eq_detcond}, see also \cref{fig_growthmode}] is somewhat larger than the system size. 
Accordingly, the instability is realized here with the largest wavelength that fits in the system (cf.\ \cref{fig_disprel}), provided that $\pi/k_c\lesssim L$, i.e., the system size exceeds half of the critical wavelength [see \cref{eq_k_crit}].
This condition constrains, \textit{inter alia}, also the value of the curvature parameter $\kappa_0$ [see \cref{eq_extra_visc}], which sets the width of the shear band interface.
Simulations with $\pi/k_m \ll L$ are typically found to be unstable at late times since the nonlinear feedback mechanism leads to a singularity in the integration of the Navier-Stokes equations (see the discussion below \cref{eq_steady_solvcond}). This singularity manifests itself in a diverging viscosity and vanishing shear rate.
For sufficiently large interface widths, global mass conservation stabilizes the stationary state before the singularity is reached.

\begin{figure}[t]\centering
    \subfigure[]{\includegraphics[width=0.335\linewidth]{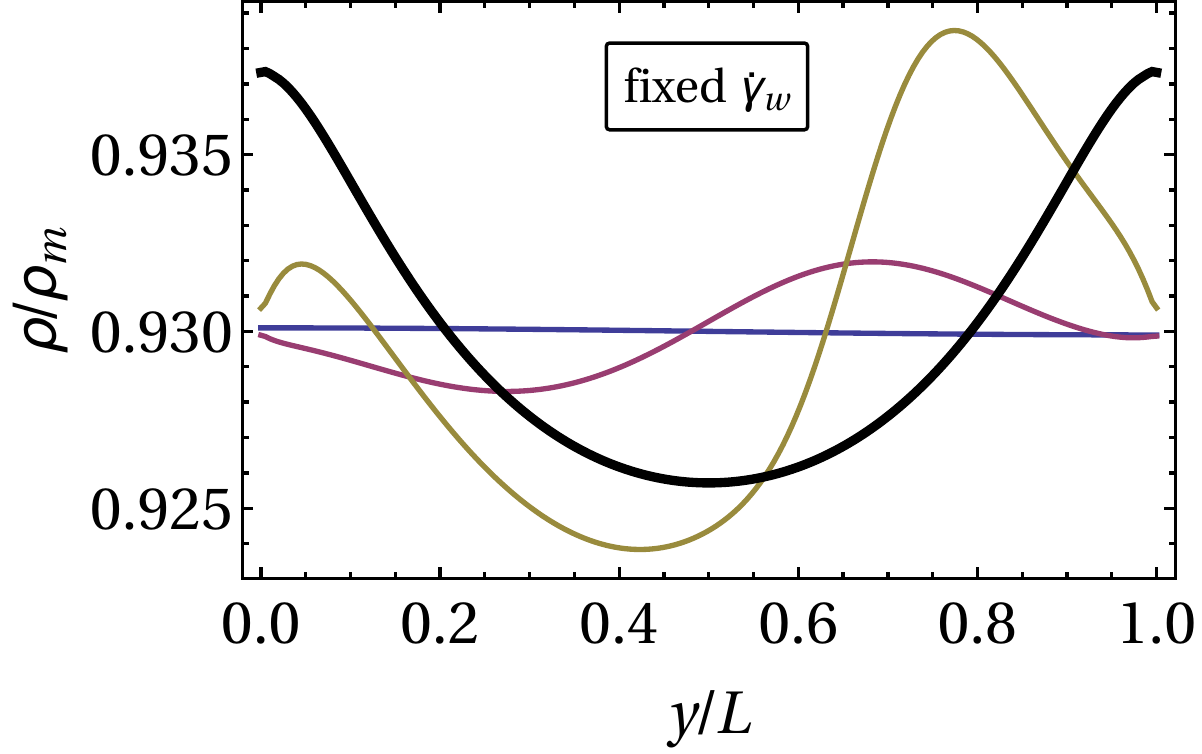} }\hfill
    \subfigure[]{\includegraphics[width=0.325\linewidth]{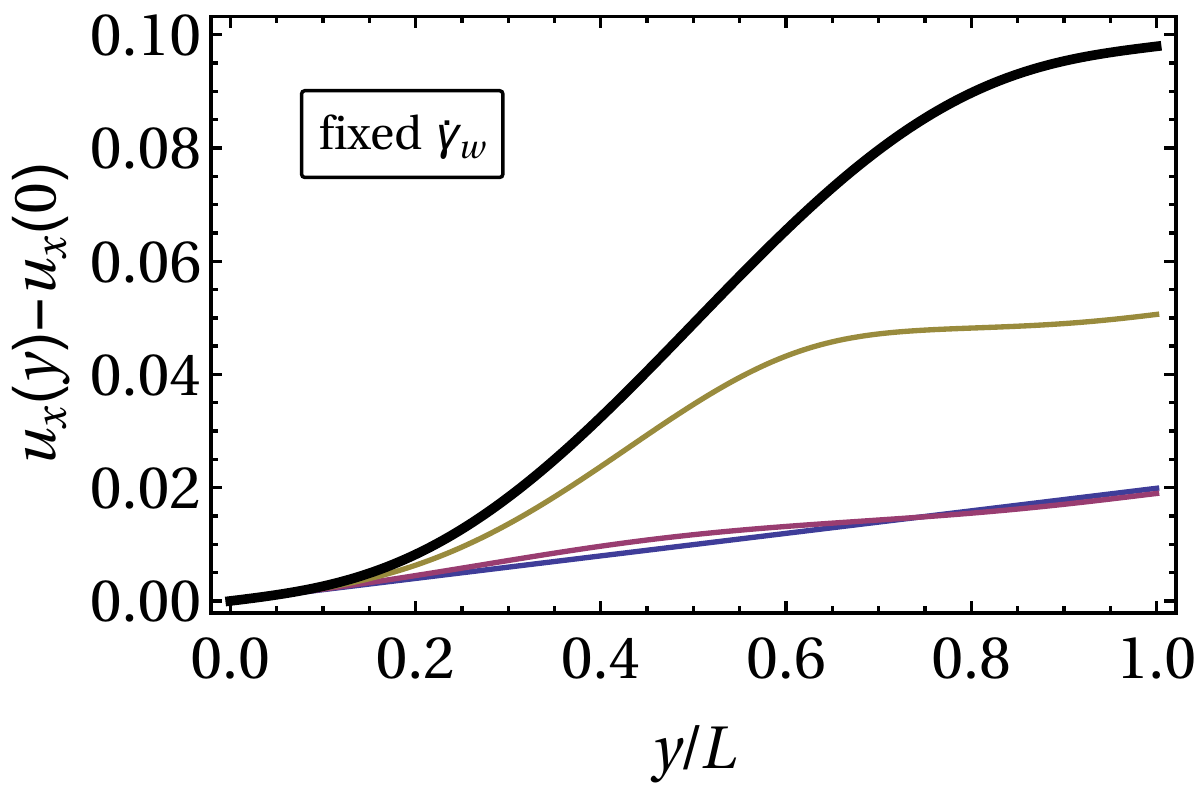} } \hfill
    \subfigure[]{\includegraphics[width=0.305\linewidth]{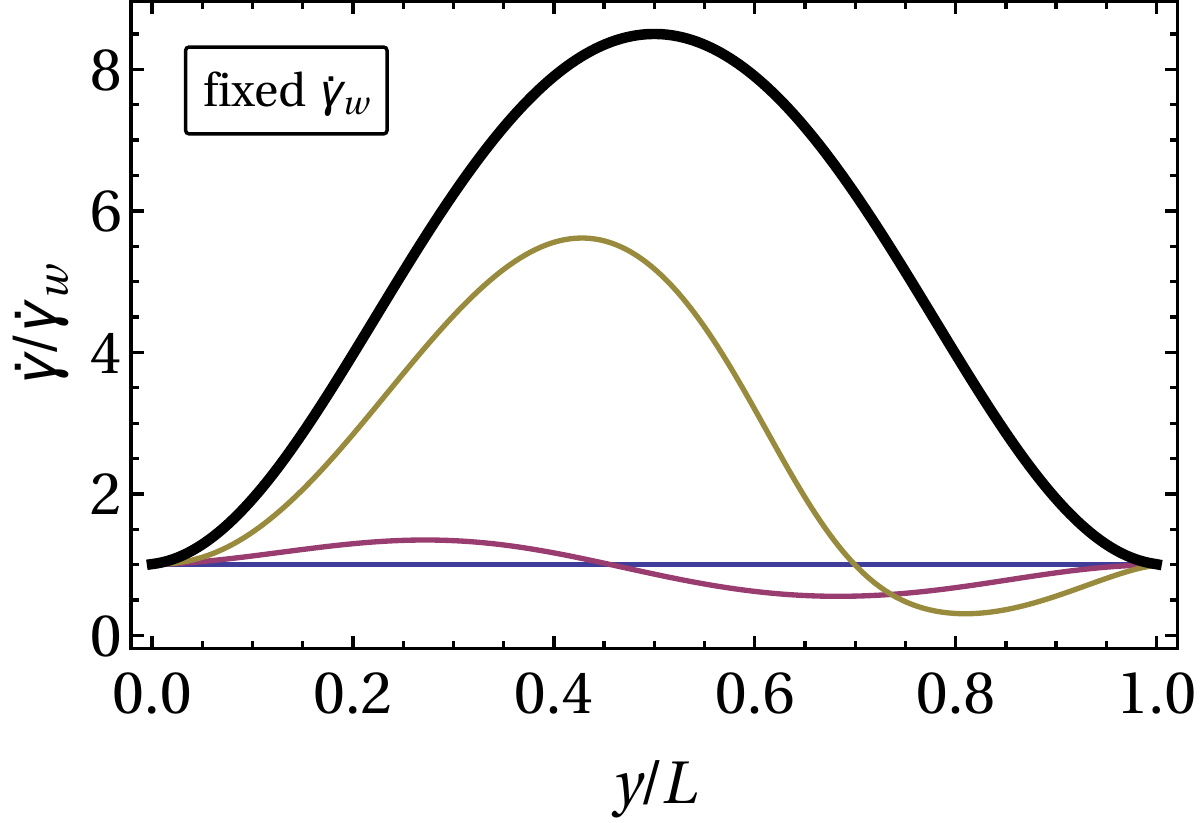} } 
    \caption{Time evolution of (a) the density, (b) the velocity, and (c) the shear rate resulting from \cref{eq_nse_red} for a fixed shear rate $\gdot_w=10^{-4}$ at both walls. The profiles shown are obtained at times $t\gdot_w =0,2.2\times 10^{-2},4.0\times 10^{-2},\infty$, where $t=\infty$ corresponds to the steady state (thick black curve) reached for $t\gdot_w\gtrsim 0.4$. In the steady-state, the local shear rate at each of the walls results as $\gdot\simeq 1.0\times 10^{-4}$. The initial growth rate of the maximally unstable mode is given by $\omega_m\simeq 0.078$ [see \cref{eq_detcond}]. Parameters $L=200\Delta h$, $\kappa_0 =100$, and $\rho_0 =0.93\rho_m$ are used.}
    \label{fig_dyn_fixedGam}
\end{figure}

\begin{figure}[t]\centering
    \subfigure[]{\includegraphics[width=0.33\linewidth]{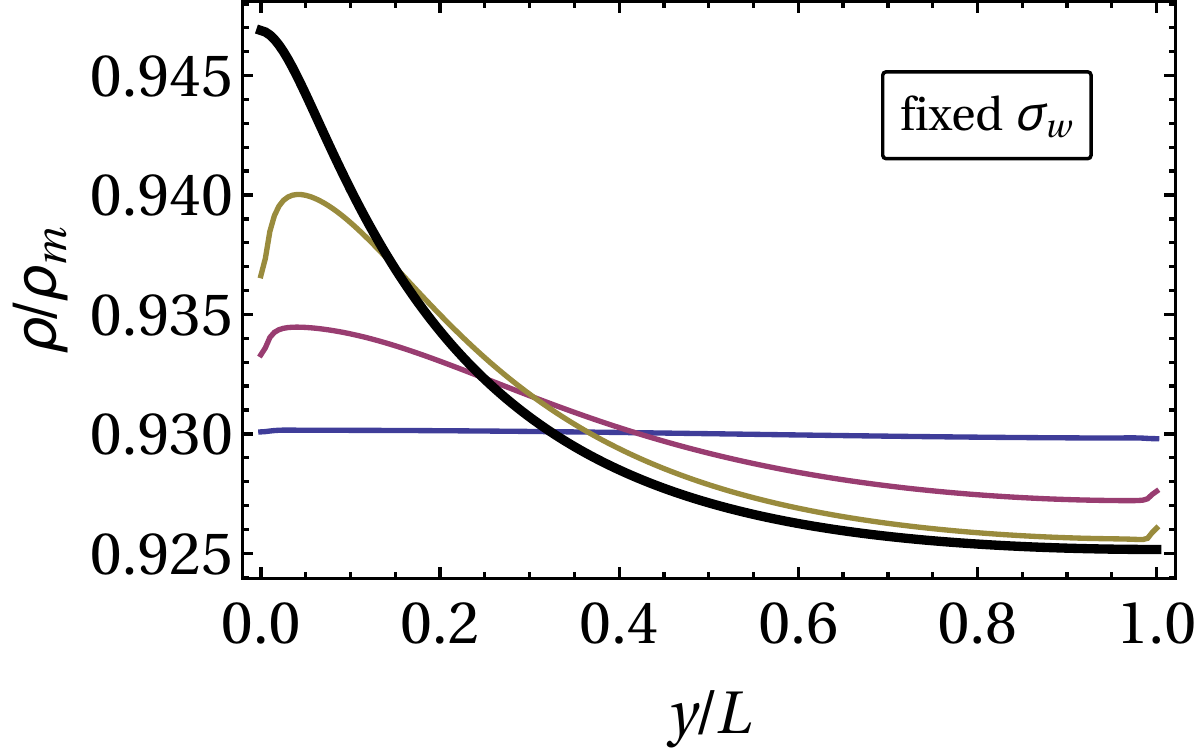} }\hfill
    \subfigure[]{\includegraphics[width=0.32\linewidth]{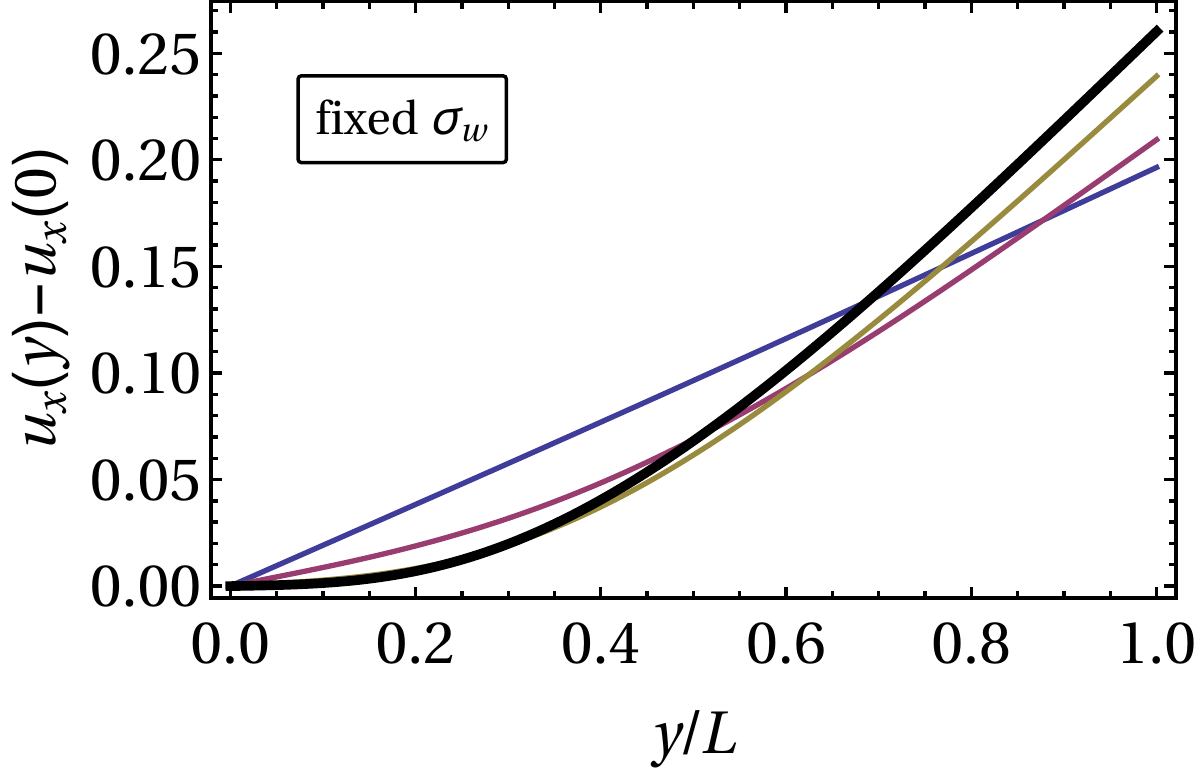} } \hfill
    \subfigure[]{\includegraphics[width=0.32\linewidth]{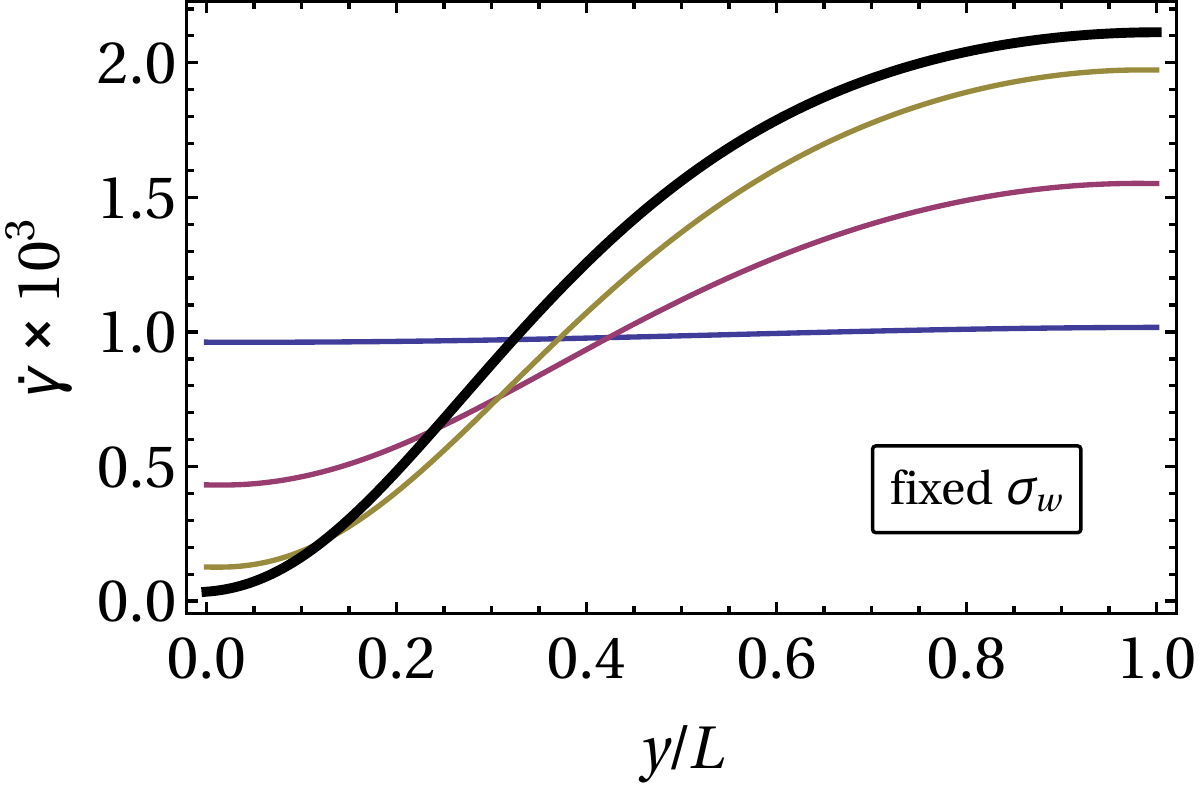} } 
    \caption{Time evolution of (a) the density, (b) the velocity, and (c) the shear rate resulting from \cref{eq_nse_red} for a fixed wall stress $\sigma_w$. The profiles shown are obtained at times $t \gdot_0 = 0.07, 0.42, 0.48,\infty$, where $t=\infty$ corresponds to the steady state (thick black curve) reached for $t\gdot_0 \gtrsim 10$. An effective shear rate $\gdot_0\simeq 10^{-3}$ is obtained from \cref{eq_steady_shearDensity_rel} based in the initial density $\rho_0$. 
    In the steady state, the local shear rate at the boundary at $y=0$ is obtained as $\gdot\simeq 3.3\times 10^{-5}$. The initial growth rate (corresponding to the above $\gdot_0$) of the maximally unstable mode is given by $\omega_m\simeq 0.012$ [see \cref{eq_detcond}]. Parameters $L=200\Delta h$, $\kappa_0 = 100$, $\rho_0 =0.93\rho_m$, and $\sigma\st{w}=11$ are used.}
    \label{fig_dyn_fixedSig}
\end{figure}

\Cref{fig_dyn_fixedU,fig_dyn_fixedGam,fig_dyn_fixedSig} illustrate the time evolution of the density $\rho$, flow velocity $u_x$, and local shear rate $\gdot=\pd_y u_x$ across the slit in the unstable region for various boundary conditions.
One observes that, in all cases, the system evolves from an essentially homogeneous initial state towards a steady state with inhomogeneous density and shear-rate profiles.
The steady state of $u_x$ (or, correspondingly, the shear rate) is typically reached within a time scale $1/\gdot_0$ determined by the average shear rate $\gdot_0$. The latter is given by $(u_x(L)-u_x(0))/L=\gdot\st{av}$ in the case where a fixed wall velocity is imposed and by $\gdot_w$ in the case where a fixed wall shear rate is used.

At late times, the evolution slows down due to the slow transport of mass towards the boundaries.
This effect is particularly pronounced in the case of a fixed wall velocity (\cref{fig_dyn_fixedU}), where the shear rate profile is essentially fully developed at times $t\gdot\st{av}\gtrsim 1$, while the density at the left wall reaches the steady state only for times $t\gdot\st{av}\gtrsim \Ocal(10^3)$. One observes that the time evolution is fastest if a fixed wall-shear rate is imposed. The broken left-right symmetry with respect to the walls in \cref{fig_dyn_fixedU,fig_dyn_fixedSig} is a direct consequence of the asymmetry of the initial configuration. In fact, using an initial sinusoidal density profile with a maximum in the right half of the system leads to spatially mirrored evolution.

The density dynamics is generally overdamped, which is expected based on an analysis of the linear equations in \cref{eq_lin_nse}:
for typical values $\rho_0$, $\gdot_0$ of the density and shear rate in the unstable regime and for wavenumbers $k\sim \Ocal(10^{-2})$ (in units of $\psize$), one finds for the present constitutive model [\cref{eq_constitut_glass}] that the viscous term $\eta\pd_y^2$ dominates over the restoring force $\Pi_\rho \pd_y^2$ in the sound-wave equation [\cref{eq_soundwave}].

\begin{figure}[t]\centering
    \subfigure[]{\includegraphics[width=0.4\linewidth]{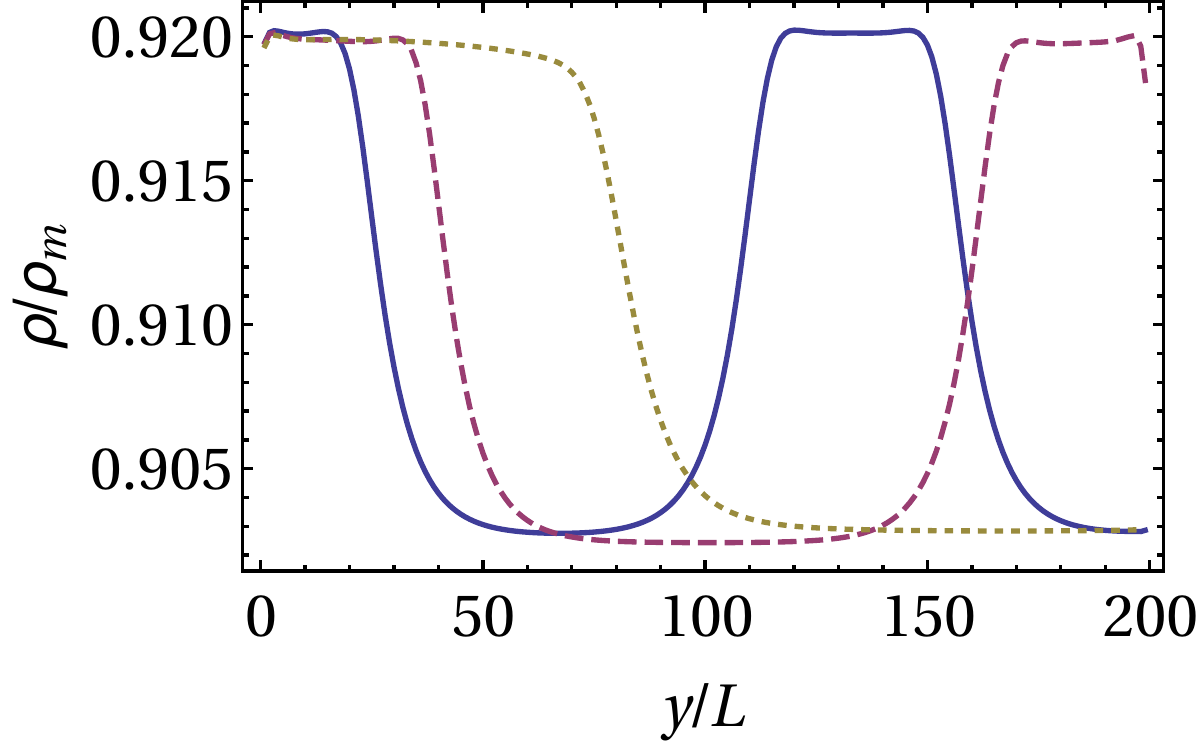} }\qquad
    \subfigure[]{\includegraphics[width=0.38\linewidth]{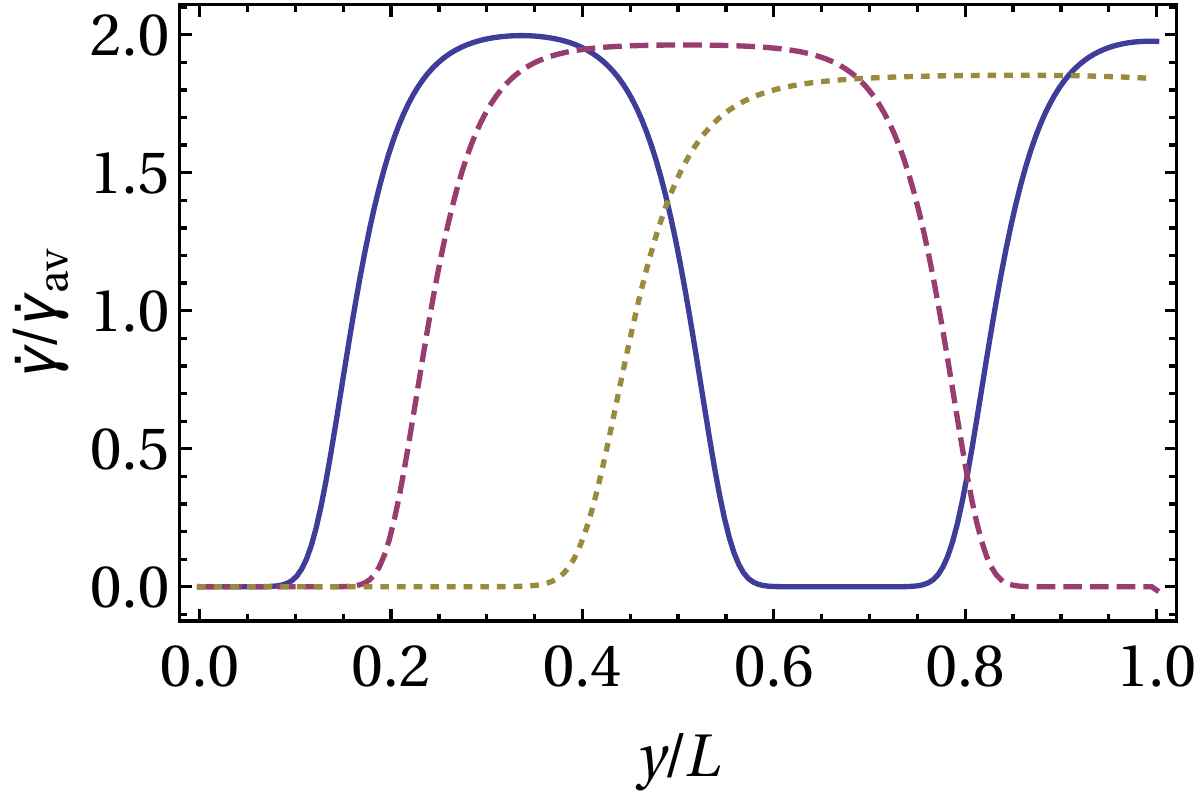} } 
    \caption{Multiple shear bands occur for sufficiently small values of the shear-curvature viscosity $\kappa_0$, which determines the effective interface width [see \cref{eq_extra_visc}]. Panel (a) shows the density profile and panel (b) the shear rate profile at times $t\gdot\st{av}\sim \Ocal(0.1)$ in the case of a fixed wall-velocity corresponding to an average shear rate $\gdot\st{av}\simeq 2\times 10^{-4}$. The initial density profile is given by $\rho(y,0)\propto \cos(n \pi y/L)$, with $n=4,2,1$ for the solid, dashed, and dotted curve, respectively. A value $\kappa_0=2$ is used, while all other parameters are the same as in \cref{fig_dyn_fixedU}. We emphasize here that, regardless of the initial structure, multiple bands are not observed for larger values of $\kappa$, such as those used in \crefrange{fig_dyn_fixedU}{fig_dyn_fixedSig}.}
    \label{fig_kappa_effect}
\end{figure}

An interesting question here regards the existence of multiple shear bands. In order to obtain such a structure, we decrease the shear-curvature viscosity $\kappa$, whereby, as stated above [see \cref{eq_extra_visc}], the width of the interface between regions of low and high shear rates is reduced. As shown in Fig.~\ref{fig_kappa_effect}, where we used a value of $\kappa_0=2$ (in dimensionless units), multiple shear bands are indeed observed in our model.
The figure also highlights the important role of the initial perturbations for the formation of a multi-banded structure, since the number of nodes of the band directly depends on the period of the initial sinusoidal profile. The shear-curvature viscosity $\kappa$, in contrast, plays a subordinate role for detailed band structure.
We emphasize that the profiles in \cref{fig_kappa_effect} are not in the steady state but correspond to times $t\gdot\st{av} \sim \Ocal(0.1)$. In fact, the presently used constitutive model does not allow us to reach the steady state in these cases due to the occurrence of an intrinsic singularity [see \cref{eq_steady_solvcond} below].

\subsection{Steady states}

In the steady state, the wall normal velocity must vanish, i.e., $u_y=0$.
It thus follows from \cref{eq_nse_red,eq_sigma_xy} that the density $\rho$ and the shear rate $\gdot$ in the steady state are determined by the equations 
\begin{subequations}\begin{align} 
0 &= \pd_y \left[ \sigma\yield(\rho) - \Pi(\rho,\gdot) \right], \label{eq_steady_st_pi}\\
0 &= \pd_y \sigma_{xy}(\rho,\gdot) = \pd_y \left[ \sigma\yield(\rho) + \eta(\rho,\gdot) \gdot - \kappa(\rho,\gdot) \pd_y^2 \gdot \right]. \label{eq_steady_st_sigmaxy}
\end{align}\label{eq_steady_st}\end{subequations}
Upon discretizing this boundary value problem using finite differences, the resulting system of nonlinear equations can be solved via Newton's method. 
In order for this scheme to converge, good initial guesses for $\rho(y)$ and $\gdot(y)$ are required, which can be obtained from the dynamical equation \eqref{eq_nse_red}, as discussed above.
Alternatively, the steady state profiles may also be directly obtained by integrating the PDEs in \cref{eq_nse_red} over a sufficiently large time, as is done in \crefrange{fig_dyn_fixedU}{fig_dyn_fixedSig}.

According to \cref{eq_steady_st}, in the steady state, the effective pressure $\Pi-\sigma\yield$ as well as the viscous stress $\sigma_{xy}$ must be constant throughout the system. 
The resulting profiles realizing these constraints are illustrated in \cref{fig_dyn_fixedU,fig_dyn_fixedGam,fig_dyn_fixedSig} (thick black curves). 
For a fixed wall velocity $u_w$ or a fixed wall stress $\sigma_w$, we obtain here spatially asymmetric steady state profiles for which the maximum density and minimum shear rate is attained close to the walls.
These profiles are qualitatively similar to the ones observed in Ref.\ \cite{jin_flow_2014}, where a fixed wall stress was considered (see also \cref{sec_AD_model}).
The spatial symmetry of the steady profile in \cref{fig_dyn_fixedGam} is a consequence of the fact that the same shear rate $\gdot_w$ is imposed at both walls.
Profile shapes similar to the ones in \cref{fig_dyn_fixedU,fig_dyn_fixedSig} result when the values of $\gdot_w$ at each wall are set accordingly (data not shown).

In the unstable region of the phase diagram, a necessary condition (which, however, is not sufficient; see below) for the development of an inhomogeneous steady state profile is the presence an initial perturbation in the system and a system size large enough such that at least one unstable mode can be accommodated. 
However, \cref{eq_steady_st} also admits constant solutions, i.e., $\rho=\rho_0$ and $\gdot=\gdot_0$. 
In this case, \cref{eq_steady_st_sigmaxy} with $\pd_y\gdot=0$ readily yields a relation between $\rho_0$, $\gdot_0$, and the stress in the system $\sigma_w$ (which arises as an integration constant and typically corresponds to the wall stress):
\beq \gdot_0 = \left[ \frac{ \frac{\sigma_w}{\sigma_0} (1-\Phi)^p - 1}{A(1-\Phi)^n} \right]^{1/n},
\label{eq_steady_shearDensity_rel}\eeq 
where $\Phi=\rho_0/\rho_m$ and we used \cref{eq_constitut_glass}.
Alternatively, \cref{eq_steady_st_pi} provides a relation between $\rho_0$, $\gdot_0$, and the system (wall) pressure. If, instead of $\sigma_w$, the wall shear rate $\gdot_w=\gdot_0$ or the wall velocity $u_w$ (implying $\gdot_0=u_w/L$) are prescribed, \cref{eq_steady_shearDensity_rel} represents a family of solutions for $\rho_0$ with the integration constant $\sigma_w$ as adjustable parameter.

\begin{figure}[t]\centering
    \includegraphics[width=0.35\linewidth]{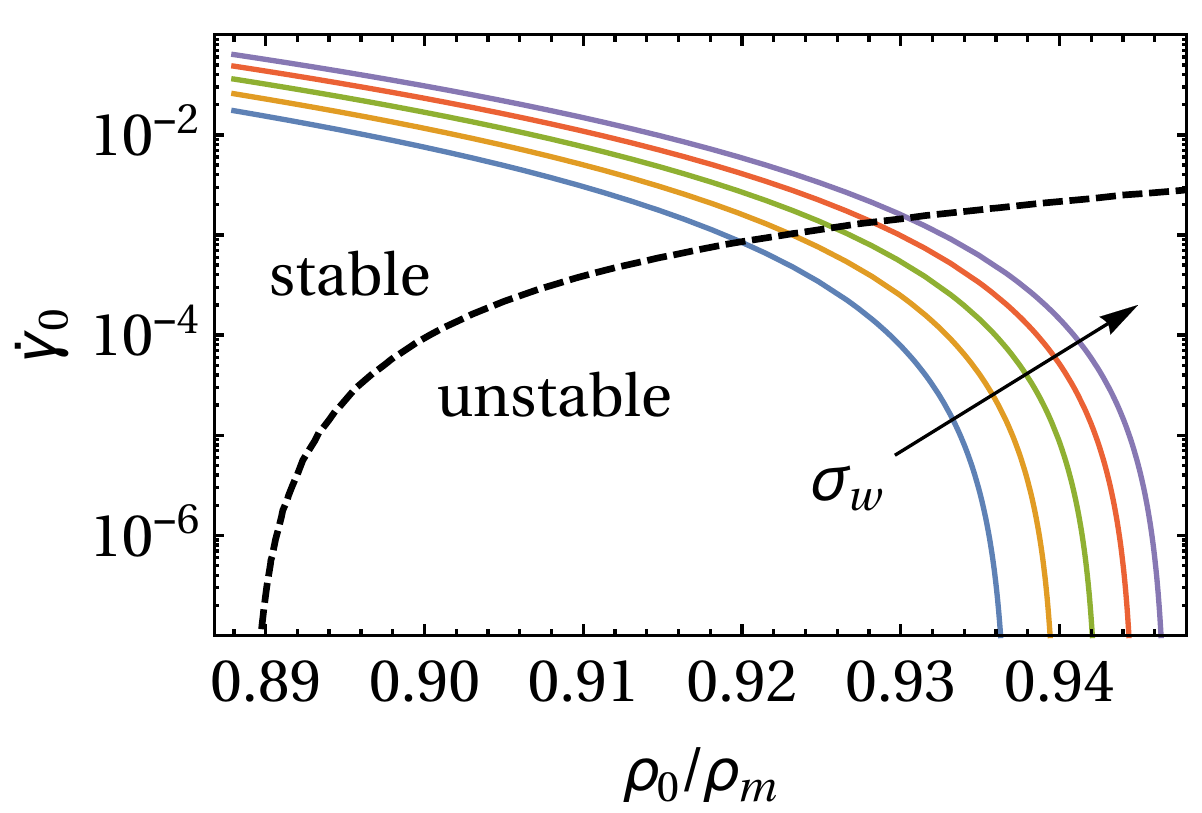} 
    \caption{Relation between shear rate $\gdot_0$ and density $\rho_0$ in a homogeneous system, as provided by \cref{eq_steady_shearDensity_rel}, for various values of the system stress $\sigma_w$ (increasing in the direction of the arrow from $\sigma_w=8$ to $12$ in steps of 1). \Cref{eq_steady_shearDensity_rel} is well-defined only if condition \eqref{eq_steady_solvcond} holds, i.e., for sufficiently small densities. The maximum possible density corresponds in the plot to the ($\sigma_w$-dependent) location where $\gdot_0\to 0$. The dashed black curve represents the threshold of the SDC instability, $\Bcal_1=0$ [see \cref{eq_stab_bound1} as well as \cref{fig_stabdiag}].}
    \label{fig_steady_wallstr}
\end{figure}

\Cref{eq_steady_shearDensity_rel} is illustrated in \cref{fig_steady_wallstr}.
A solution of \cref{eq_steady_shearDensity_rel} exists provided that the numerator on the r.h.s.\ is positive, i.e., for
\beq \Phi\leq 1-(\sigma_0/\sigma_w)^{1/p},
\label{eq_steady_solvcond}\eeq 
or, equivalently, if the yield stress remains below the system stress,
\beq \sigma\yield(\Phi) \leq \sigma_w.
\label{eq_steady_solvcond2}
\eeq 
For $\sigma\yield > \sigma_w$, instead, the system becomes increasingly more rigid and thus ceases to flow.
This singular behavior is indeed reflected in the solution of the fluid dynamical equations [\cref{eq_nse_red}] in the SDC-unstable region. In order to gain a heuristic understanding of this singularity, let us assume that, close to the inhomogeneous steady state, the density and shear rate profiles in the fluid consists of large nearly flat portions.
In each of these regions then \cref{eq_steady_shearDensity_rel} and thus \cref{eq_steady_solvcond} approximately hold, with $\sigma_w$ being the corresponding local stress.
In order to trigger the SDC instability, here typical values of $\sigma_w\simeq \Ocal(10)$ are required \footnote{For fixed wall velocity or fixed wall shear rate boundary conditions,  the value of $\sigma_w$ is found to change only weakly during the time evolution. Accordingly, $\sigma_w$ is essentially set by the chosen initial density and shear rate profiles.}, for which \cref{eq_steady_solvcond} implies $\Phi \sim \Ocal(0.94)$ as an upper limit for the density (see \cref{fig_steady_wallstr}). As indicated in \cref{fig_fluctdir}, once the system is unstable, some part of it evolves towards smaller shear rates and larger densities.
In small systems, this growth is eventually limited by global mass conservation and the fact that the interface between low and high density regions must keep a certain width [see discussion after \cref{eq_extra_visc}]. 
In contrast, for systems much larger than the interface width, the limit for $\Phi$ implied by \cref{eq_steady_solvcond} can be easily exceeded by means of mass transport, leading to $\sigma\yield>\sigma_w$ and thus causing the solution of \cref{eq_nse_red} to become singular. 
The singularity, in particular, makes an observation of stationary inhomogeneous profiles with multiple shear bands rather difficult. Nevertheless, as shown in \cref{fig_kappa_effect}, we do observe a multi-band structure up to the instant of the numerical singularity. Based on these findings and the above detailed analysis, we anticipate that a slightly modified version of the present constitutive model with a tamed singularity would exhibit stable multiple shear bands.

Physically, the singularity can be understood as ``freezing''---a behavior which is a hallmark of yield stress fluids upon increasing the density or decreasing the shear rate. Accordingly, a possibility to avoid this singularity would be to limit the growth of the yield stress and the viscosities in the constitutive equations \eqref{eq_constitut_glass} and \eqref{eq_constitut_press}. An adequate study of this issue is an interesting topic for future work.

\section{Summary}

This study addresses the issue of flow heterogeneity, often observed in the glassy state of matter under externally imposed shear. We focus on the limit of gently sheared dense \emph{single-component} fluids, such as colloidal hard sphere glasses, where hydrodynamic interactions are negligible. Therefore, the effect of solvent is ignored in this study. Instead of a coupling to the concentration field as in the original theory of shear-concentration coupling \cite{schmitt_shear-induced_1995}, in the present case, fluctuations of the velocity field are coupled to fluctuations of the fluid density. 
Analogously to a standard density-dependent thermodynamic pressure, here, a shear-rate dependent pressure drives a transverse flow away from regions of high shear rate, thus further lowering the viscosity in that region.
Together with a strongly non-Newtonian viscous stress, this gives rise to a feed-back mechanism which ultimately determines the borders of flow stability. A detailed analysis of the resulting cubic dispersion relation reveals that the instability can occur only via a monotonic growth of fluctuations, thus excluding the possibility of an oscillatory growth mode. Notably, after an initial transient, the velocity and density fields reach a stationary profile. In this stationary state, regions of high shear rate exhibit low density and vise versa. 
For systems much larger than the characteristic width of the shear-band interface, the fluid model considered here generally develops a singularity in the unstable regime, accompanied by a vanishing shear rate and thus a divergent viscosity.
The steady states obtained here in fact all occur for system sizes comparable to the interface width, in which case they are stabilized by means of global mass conservation.

Interestingly, the expression for the stability threshold and the range of unstable wavenumbers are identical to those obtained from an analysis of the advection-diffusion equation based on the original theory of shear-concentration coupling \cite{jin_flow_2014}. The difference between the present compressible single-component fluid model and the one incorporating the coupling of flow to a concentration field is exhibited in the specific dynamics and underlying timescales, such as the expression for the fastest growing mode. 
In the compressible fluid, the density relaxes via overdamped sound waves---a transport mechanism which, in contrast to diffusion, leads to wavenumber-independent exponential relaxation in the limit of small frequencies and large wavenumbers.
The use of a compressible fluid model is supported by molecular dynamics simulations \cite{mandal_heterogeneous_2012}, which show that variations of the density are a typically observed response to fluctuations of the shear rate in single-component hard-sphere colloidal glasses.

The existence of a stationary solution seems, at first sight, to be in conflict with the time dependent behavior of the shear band observed in molecular dynamics simulations \cite{mandal_heterogeneous_2012}. A plausible interpretation here would be to invoke the coupling between velocity fluctuations and structural heterogeneity in the glassy state. As also discussed in Ref.~\cite{besseling_shear_2010}, this may lead to the formation of a locally depleted zone with a density in the stable regime and a denser packing in the remaining part of the system with enhanced instability and a corresponding temporal evolution. This is also in-line with recent reports on the strong influence of structural heterogeneity on plastic deformation in the amorphous solid state \cite{rosner_density_2014,schmidt_quantitative_2015,hassani_localized_2016}. Notably, the time scale of the shear band dynamics in MD simulations is of the order of the inverse shear rate \cite{mandal_heterogeneous_2012}. This is also the time associated with structural fluctuations, since during this time a particle moves a distance comparable to its size and the cage of nearest neighbors around it relaxes to a large extent. This stochastic effect is not included in the present deterministic model. A way to account for this would be to add a noise term into hydrodynamic equations, which is left for future work.

\section{Acknowledgments}
This work is supported by the German Research Foundation (DFG) under the project number VA 205/18-1. ICAMS acknowledges funding from its industrial sponsors, the state of North-Rhine Westphalia and the European Commission in the framework of the European Regional Development Fund (ERDF).

\appendix

\section{Transport mechanism}
\label{app_transport}

Here, we provide further insights into the transport mechanism of the compressible fluid model, as compared to the diffusive transport model studied in the original SCC theory \cite{schmitt_shear-induced_1995,jin_flow_2014}.
In order to focus on the essential aspects, we consider linearized dynamics.
In both models, shear rate fluctuations are governed by \cref{eq_shear_rt_diffusion}, i.e., a diffusion equation with a coupling to density fluctuations. (The last term in \cref{eq_shear_rt_diffusion} is typically small in our case and absent in the model of Ref.\ \cite{jin_flow_2014}.)
However, instead of following a diffusion equation as in Ref.\ \cite{jin_flow_2014,schmitt_shear-induced_1995}, density fluctuations $\delta\rho$ in the compressible fluid are governed by the generalized sound-wave equation \eqref{eq_soundwave}. In the glassy state, the viscosities are large and the dynamics is thus strongly overdamped, such that the term $\pd_t^2\delta\rho$ can be neglected in \cref{eq_soundwave}, resulting in 
\beq 0 \simeq \frac{\bvisc}{\rho_0} \left(\eta - \kappa\pd_y^2\right) \pd_y^2 \pd_t\delta\rho + \Pi_\gdot \pd_y^2\delta\gdot + \Pi_\rho\pd_y^2\delta\rho.
\label{eq_soundwave_overdmp0}\eeq 
Noting that the kinetic coefficients are constants here and focusing on large wavelengths, where the term involving $\kappa$ can be disregarded, \cref{eq_soundwave_overdmp0} reduces, after two integrations over $y$, to
\beq \pd_t\delta \rho = - A \delta\rho - B \delta\gdot + c + d y,
\label{eq_soundwave_overdmp}\eeq 
with $A\equiv \sfrac{\rho_0 \Pi_\rho}{b\eta}$, $B\equiv \sfrac{\rho_0 \Pi_\gdot}{b \eta}$, and integration constants $c$ and $d$.
In order to have $\delta \rho\simeq 0$ at the boundaries of the domain, we set $d=0$, such that the solution of \cref{eq_soundwave_overdmp} with initial condition $\rho(y,0) = \rho\st{in}(y)$ is obtained as
\beq \delta \rho(y,t) = \frac{c}{A}\left(1-e^{-A t}\right) + e^{-A t} \rho\st{in}(y) - e^{-A t} \int_0^t \d s\, e^{A s} B \delta \gdot(y,s).
\label{eq_sound_sol}\eeq 
Due to the neglect of the term $\pd_t^2 \delta\rho$, \cref{eq_soundwave_overdmp} does not conserve mass. 
The effect of global mass conservation can be mimicked in \cref{eq_sound_sol} by setting $c=A\int \d y' \rho\st{in}(y')$, which follows from requiring $0=\int \d y' \delta\rho(y',t=0)$.
\Cref{eq_sound_sol} shows that, in the overdamped limit and for large wavelengths and small frequencies, density fluctuations essentially relax exponentially in the compressible fluid (cf.\ Ref.~\cite{gross_critical_2012}).
However, in contrast to diffusive relaxation, which is also exponential at late times, the relaxation rate is here independent of the wavenumber. 
Furthermore, according to \cref{eq_sound_sol}, a positive shear rate fluctuation gives rise to a reduction of the local density. This behavior is an essential mechanism of the SDC instability.

\section{Diffusive transport model}
\label{sec_AD_model}
Here, we compare our results obtained in \cref{sec_nonlin} to the SCC model studied in in Ref.\ \cite{jin_flow_2014}, which is based on the advection-diffusion equation for the concentration given in \cref{eq_advdiff}.
The flow velocity $\uv$ is assumed to relax much faster than the density, such that the shear-rate is essentially enslaved to the density evolution. Furthermore, also advective transport is neglected, such that the SCC model as considered in Ref.\ \cite{jin_flow_2014} effective reduces to a purely diffusive transport model:
\begin{subequations}\begin{align}
\pd_t \rho  &= \pd_y^2 \Pi(\rho,\gdot), \label{eq_dyn_AD1} \\
0 &= \pd_y \sigma_{xy}(\rho,\gdot), \label{eq_dyn_AD2}
\end{align}\label{eq_dyn_AD}\end{subequations}
where $\sigma_{xy}$ is given by \cref{eq_sigma_xy}.
In writing \cref{eq_dyn_AD1}, we used the fact that, for the constitutive model in \cref{eq_constitut_press}, the effective diffusivity $D\eff$ and the shear-gradient coefficient $\xi$ defined in Ref.\ \cite{jin_flow_2014} derive from the pressure $\Pi(\rho,\gdot)$ via $D\eff \equiv \pd_\rho \Pi$ and $\xi \equiv \pd_\gdot \Pi$, such that $\pd_y^2\Pi = \pd_y\left[ D\eff \pd_y\rho + \xi\pd_y \gdot \right]$.

Instead of the Couette geometry considered in Ref.\ \cite{jin_flow_2014}, we study here the time evolution of \cref{eq_dyn_AD} for a planar shear flow (see \cref{fig_setup}).
As in the main text, we impose either a fixed wall velocity $u_w$, a fixed wall shear rate $\gdot_w$, or a fixed wall stress $\sigma_w$ at the boundaries.
For the first two cases, instead of \cref{eq_dyn_AD2}, we solve the full time-dependent equation \cref{eq_nse_simp_jx} for the velocity $u_x$,
\beq \pd_t (\rho u_x) = \pd_y \sigma_{xy}(\rho, \gdot),
\label{eq_dyn_AD_ux}\eeq 
with $\gdot = \pd_y u_x$.
We generally impose a vanishing pressure gradient $\pd_y\Pi=0$ at the boundaries, which ensures global mass conservation for the dynamics described by \cref{eq_dyn_AD1}.

\begin{figure}[t]\centering
    \subfigure[]{\includegraphics[width=0.35\linewidth]{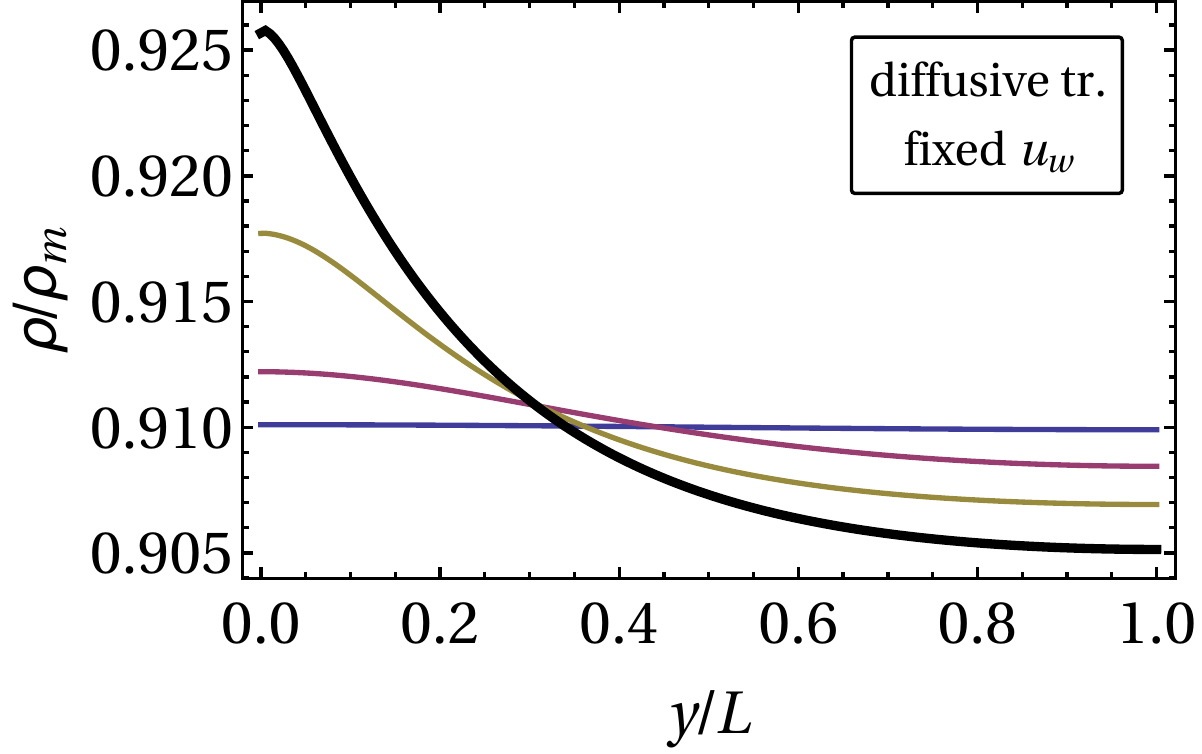} }\qquad
    \subfigure[]{\includegraphics[width=0.33\linewidth]{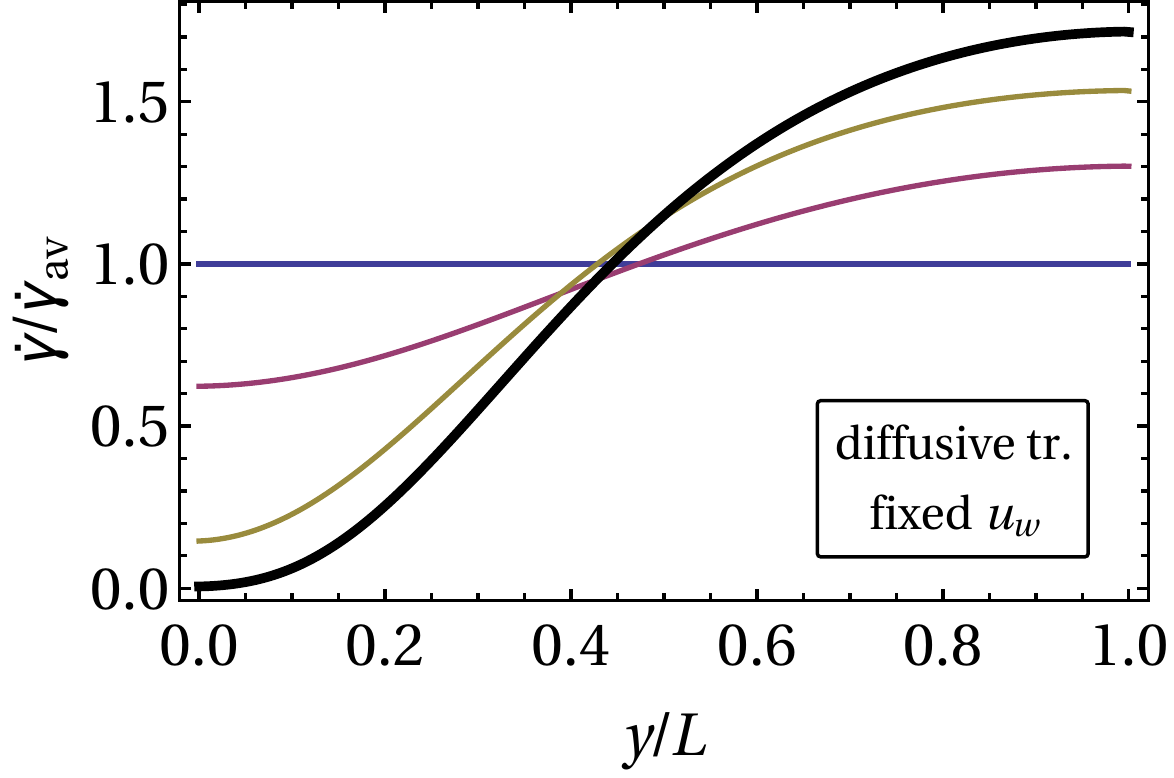} } 
    \caption{Diffusive transport model [\cref{eq_dyn_AD1,eq_dyn_AD_ux}]: time evolution of (a) the density and (b) the shear rate for a fixed wall velocity $u_w=u_x(L)-u_x(0)$ corresponding to an average shear rate $\gdot\st{av}=u_w/L=2\times 10^{-4}$. The profiles shown are obtained at times $t\gdot\st{av}=0, 0.38, 0.4,\infty$,  where $t=\infty$ corresponds to the steady state (thick black curve) reached for $t\gdot\st{av} \gtrsim 0.8$. The shear rate at the wall is obtained as $\gdot\simeq 1.1\times 10^{-6}$. The growth rate of the maximally unstable mode is given by $\omega_m\simeq 7.8\times 10^{-3}$. Parameters $L=200\Delta h$, $\kappa_0=100$, and $\rho\st{av}=0.91\rho_m$ are used.}
    \label{fig_dynAD_fixedUw}
\end{figure}

\begin{figure}[t]\centering
    \subfigure[]{\includegraphics[width=0.35\linewidth]{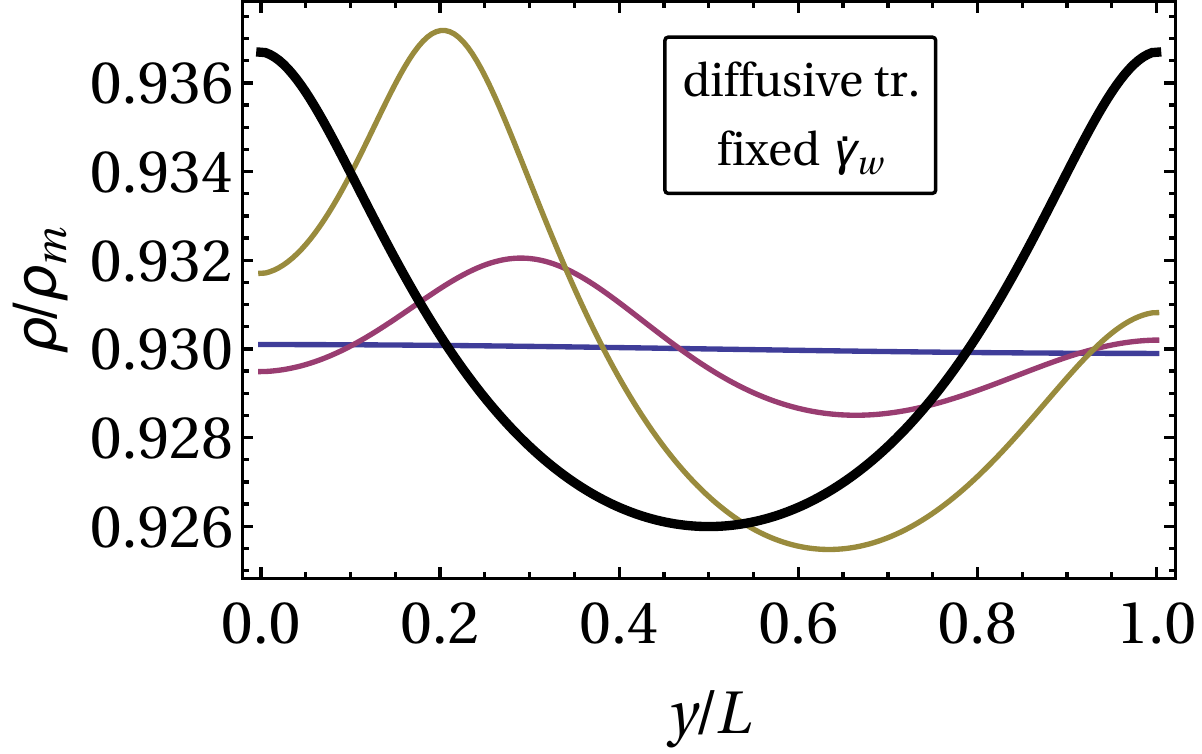} }\qquad
    \subfigure[]{\includegraphics[width=0.32\linewidth]{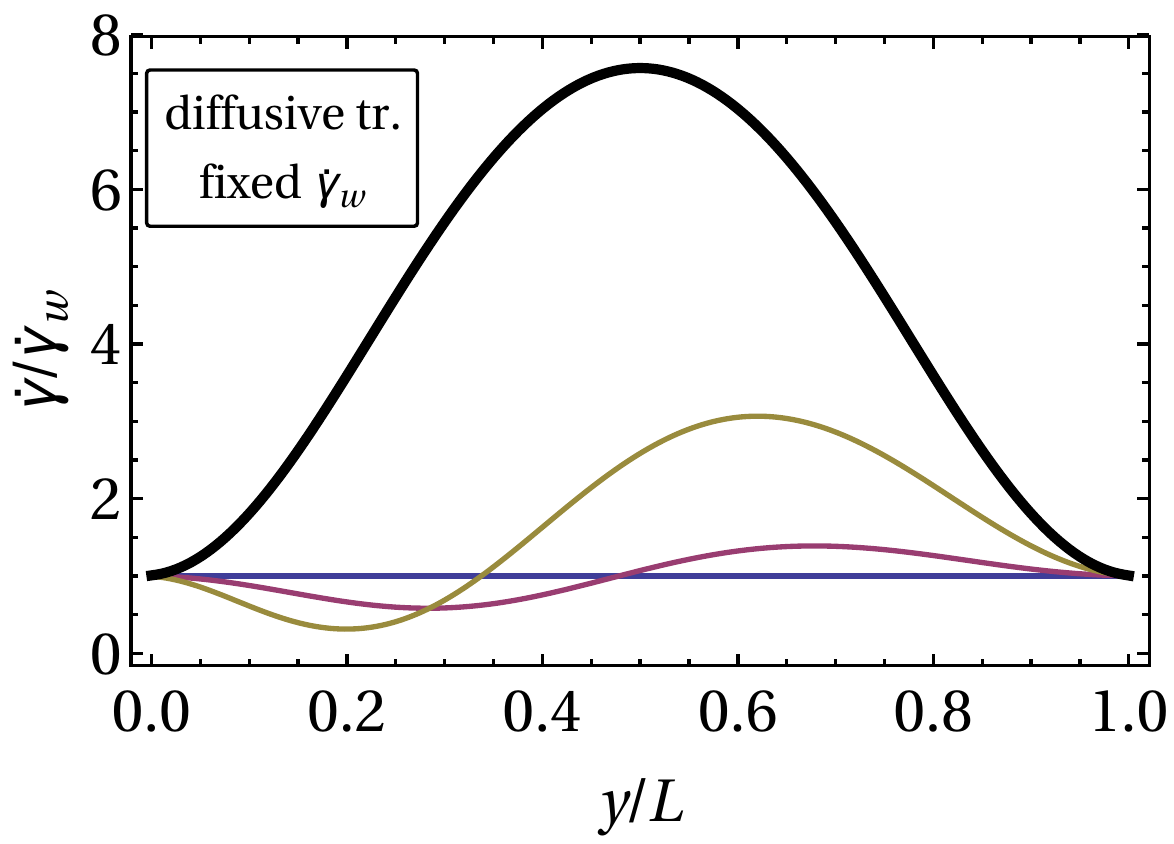} } 
    \caption{Diffusive transport model [\cref{eq_dyn_AD1,eq_dyn_AD_ux}]: Time evolution of (a) the density and (b) the shear rate resulting from \cref{eq_dyn_AD} for a fixed shear rate $\gdot_w=10^{-4}$ at both walls. The profiles shown are obtained at times $t\gdot_w=0,3.5\times 10^{-4}, 5.9\times 10^{-4}, \infty$, where $t=\infty$ corresponds to the steady state (thick black curve) reached for $t\gdot_w \gtrsim 10^{-2}$. The growth rate of the maximally unstable mode is given by $\omega_m\simeq 0.078$. Parameters $L=200\Delta h$, $\kappa_0=100$, and $\rho\st{av}=0.93\rho_m$ are used}
    \label{fig_dynAD_fixedGamw}
\end{figure}

\begin{figure}[t]\centering
    \subfigure[]{\includegraphics[width=0.35\linewidth]{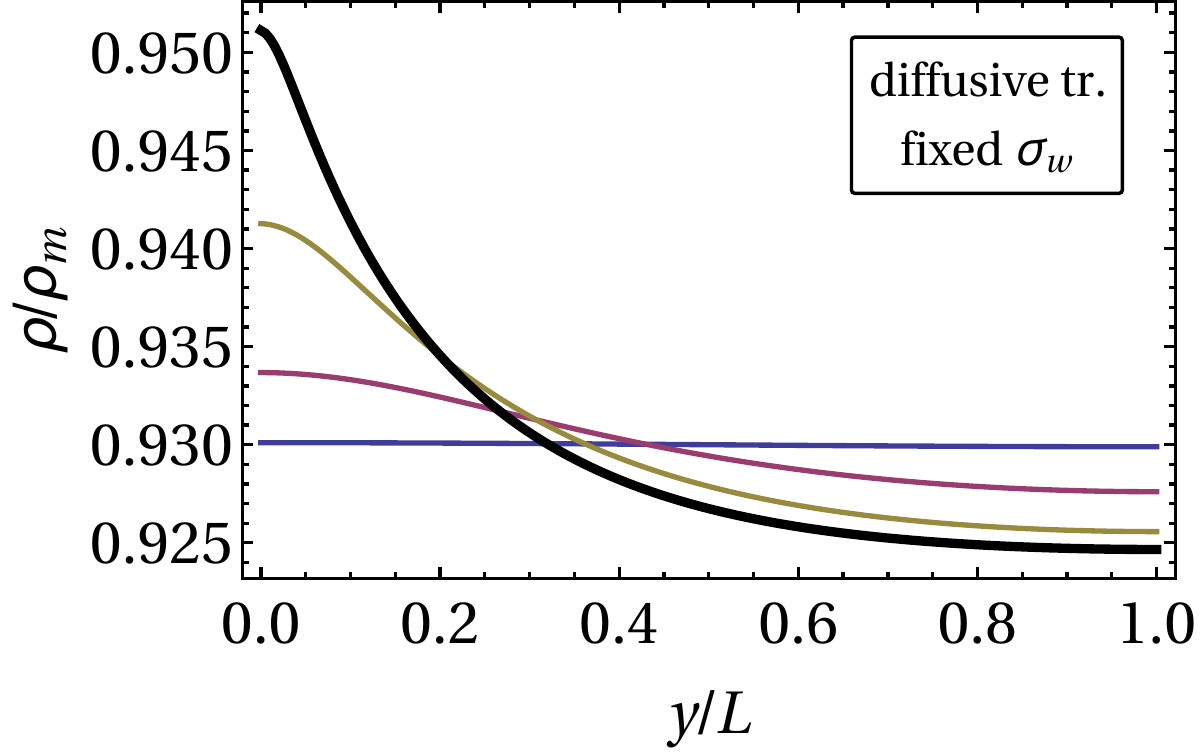} }\qquad
    \subfigure[]{\includegraphics[width=0.34\linewidth]{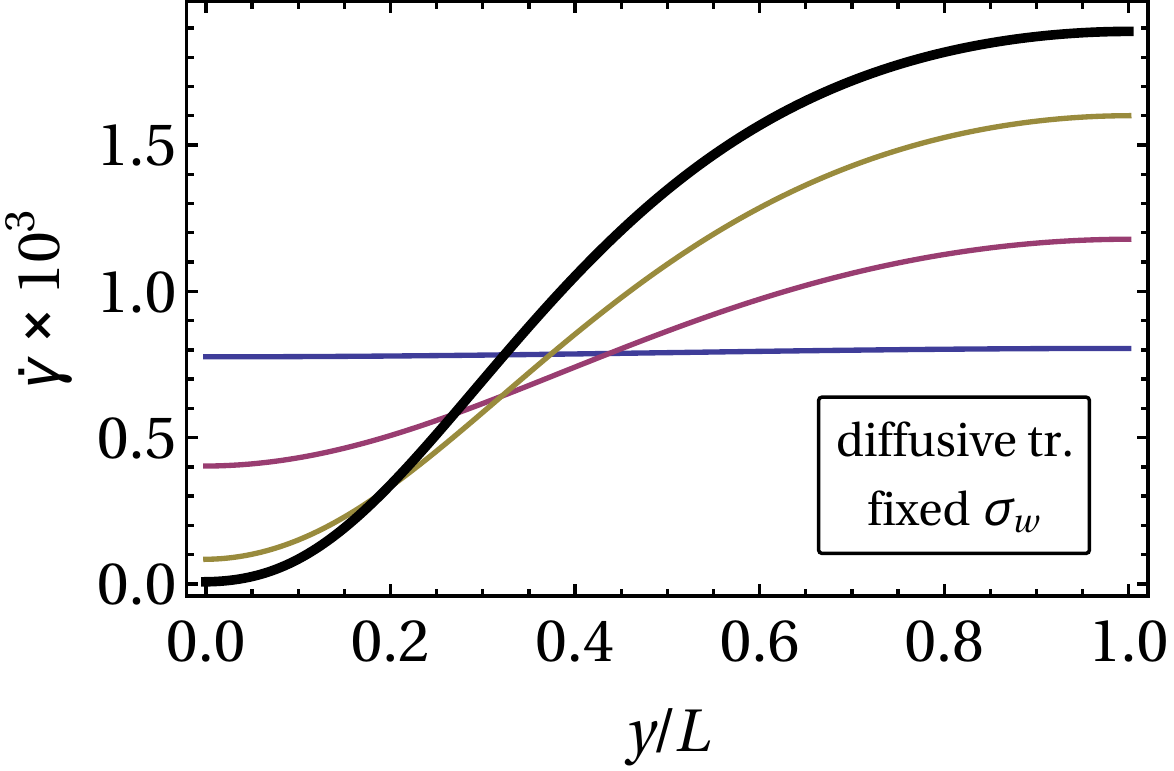} } 
    \caption{Diffusive transport model [\cref{eq_dyn_AD}]: Time evolution of (a) the density and (b) the shear rate resulting from \cref{eq_dyn_AD} for a fixed wall stress $\sigma_w$.  The profiles shown are obtained at times $t\gdot_0=0, 0.06,  0.07, \infty$,  where $t=\infty$ corresponds to the steady state (thick black curve) reached for $t\gdot_0 \gtrsim 0.1$. The effective shear rate $\gdot_0\simeq 8\times 10^{-4}$ is identified from \cref{eq_steady_shearDensity_rel} using the initial density $\rho_0$. The growth rate of the maximally unstable mode is given by $\omega_m\simeq 0.028$. Parameters $L=200\Delta h$, $\kappa_0=100$, $\rho\st{av}=0.93\rho_m$, and $\sigma\st{w}=10.6$ are used.}
    \label{fig_dynAD_fixedSig}
\end{figure}

\Crefrange{fig_dynAD_fixedUw}{fig_dynAD_fixedSig} illustrate the time evolution of the density and the shear rate in an unstable system for various boundary conditions.
In all cases, the density profile is initialized with a sinusoidal modulation in order to trigger the initial instability.
In general, the steady state is reached significantly faster for diffusive dynamics than for the compressible fluid model described by \cref{eq_nse_red}.
Both for fixed $u_w$ and $\sigma_w$, the steady state is typically reached for strains $t\gdot\st{av}\sim \Ocal(1)$ and $\sim \Ocal(0.1)$, respectively, while, for fixed $\gdot_w$, instead, the growth rate of the maximally unstable mode $\omega_m$ provides a better estimate of the dynamical time scale than the strain.
(The value of the time scale inferred from simulation depends somewhat on the chosen initial configuration.)
The larger steady-state time scale in the compressible fluid model is predominantly caused by the slow transport of mass towards the wall at the late stages of the evolution.
In fact, apart from this difference, the spatio-temporal evolution in both the compressible and the diffusive transport model are qualitatively similar (cf.\ \crefrange{fig_dyn_fixedU}{fig_dyn_fixedSig}).

\begin{figure}[t]\centering
    \subfigure[]{\includegraphics[width=0.35\linewidth]{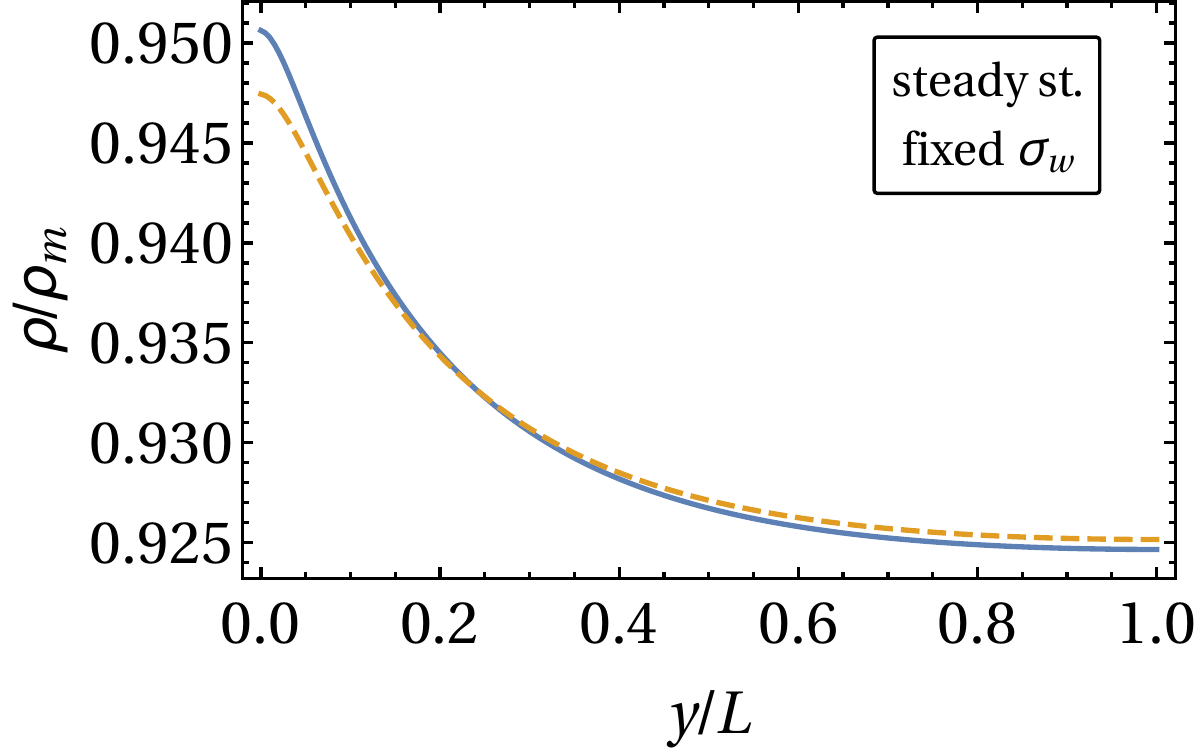} }\qquad
    \subfigure[]{\includegraphics[width=0.34\linewidth]{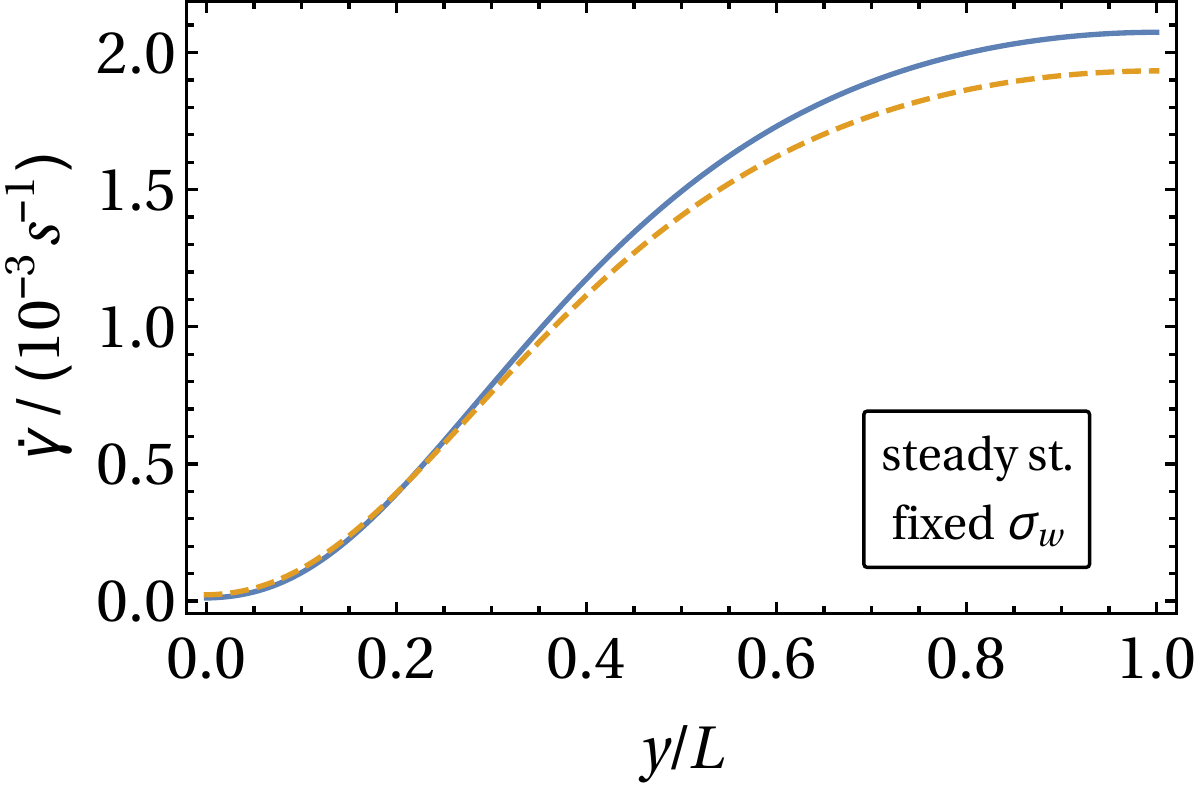} } 
    \caption{Comparison of the typical steady states obtained from \cref{eq_steady_st} (solid lines) and \cref{eq_dyn_AD} (dashed lines) for a fixed wall stress $\sigma_w$. Parameters $\sigma\st{w}=10.8$, $L=200\Delta h$, $\kappa_0=100$, and $\rho\st{av}=0.93\rho_m$ are used.}
    \label{fig_stdy_comp_sigW}
\end{figure}

Since we impose a vanishing pressure gradient at the boundaries when solving \cref{eq_dyn_AD1}, the \emph{steady states} resulting from \cref{eq_steady_st,eq_dyn_AD} are characterized by $\Pi$ and $\sigma_{xy}$ being constant throughout the system.
Thus the only difference between the steady states of two models stems from the presence of the yield stress in \cref{eq_steady_st_pi}.
Accordingly, the steady state-profiles obtained from \cref{eq_steady_st} and \cref{eq_dyn_AD} are very similar, which is illustrated in \cref{fig_stdy_comp_sigW} for the case of fixed wall stress boundary conditions.
It is therefore not surprising that the time evolution, when starting from the same initial conditions, is similar in the two models.

%

\end{document}